\documentclass[12pt,french,epsf,psfig]{elsart} 
\usepackage[latin1]{inputenc}
\usepackage[T1]{fontenc}
\usepackage{fancyhdr}
\usepackage{amsmath}
\usepackage{graphics}
\usepackage{subfigure}
\usepackage[final]{epsfig}
\usepackage{mathrsfs}
\usepackage{amssymb}
\usepackage{multirow}
\usepackage{rotating}
\makeatletter
\newcommand{\LyX}{L\kern-.1667em\lower.25em\hbox{Y}\kern-.125emX\spacefactor1000}
\newcommand{\eq}[1]{Eq.~\ref{eq:#1}} 
\newcommand{\fig}[1]{Fig.~\ref{fig:#1}}

\newcommand{\para}[1]{Sec.~\ref{para:#1}} 
 
\makeatother 
 
\newcommand\myCaption[1]{\small\refstepcounter{table}
   \centering\tablename\ \thetable :\ #1}
\newcommand\myKaption[1]{\small\refstepcounter{table}
   \tablename\ \thetable :\ #1}

\begin{document} 
 
\begin{frontmatter}
 
\title{Longwave radiative analysis of cloudy scattering atmospheres using a Net Exchange Formulation}

\maketitle
\begin{center}
\author[Toulouse]{V. Eymet\corauthref{corres}}
\author[Paris]{J.L. Dufresne}
\author[SB]{P. Ricchiazzi}
\author[Toulouse]{R. Fournier}
\author[Toulouse]{S. Blanco}
\corauth[corres]{Corresponding author. Present address~: Laboratoire d'Energétique, UFR-PCA, Université Paul Sabatier, 118 route de Narbonne, 31062 Toulouse Cedex, FRANCE \newline {\it E-mail address:} eymet@energetique.ups-tlse.fr  (V. Eymet)}
\address[Toulouse]{Laboratoire d'Energétique, Université Paul Sabatier, 31062 Toulouse, France}
\address[Paris]{LMD/CNRS, Université de Paris VI, 75252 Paris, France}
\address[SB]{Institute for Computational Earth System Science, University of California at Santa Barbara, Santa Barbara, California}
\end{center}

\begin{abstract}
The Net Exchange Formulation (NEF) is an alternative to the usual radiative transfer equation. It was proposed in 1967 by Green \cite{Green} for atmospheric sciences and by Hottel \cite{Hottel:67} for engineering sciences. Until now, the NEF has been used only in a very few cases for atmospheric studies. Recently we have developped a longwave radiative code based on this formulation for a GCM of the Mars planet. Here, we will present results for the Earth atmosphere, obtained with a Monte Carlo Method based on the NEF. In this method, fluxes are not addressed any more. The basic variables are the net exchange rates (NER) between each pair of atmospheric layer $(i,j)$, i.e. the radiative power emitted by $i$ and absorbed by $j$ minus the radiative power emitted by $j$ and absorbed by $i$. The graphical representation of the NER matrix highlights the radiative exchanges that dominate the radiative budget of the atmosphere and allows one to have a very good insight of the radiative exchanges. Results will be presented for clear sky atmospheres with Mid-Latitude Summer and Sub-Arctic Winter temperature profiles, and for the same atmospheres with three different types of clouds. The effect of scattering on longwave radiative exchanges will also be analysed.
\end{abstract}
\end{frontmatter}

\section{Introduction}

The Net Exchange Formulation (NEF) is an alternative to the usual radiative transfer formulation. It was proposed by Green \cite{Green} for atmospheric sciences and by Hottel and Sarofim \cite{Hottel:67} for engineering sciences. Joseph and Bursztyn \cite{Jose:76} attempted to use this approach to model radiative transfer in the terrestrial atmosphere. They showed that radiative net-exchanges between an atmospheric layer and the boundaries (space and ground) are dominant and that net-exchanges with the rest of the atmosphere contribute to around 15$\%$ of the total energy budget, but they encountered some numerical difficulties. Stephens et al. \cite{Stephens:84} also presented this formalism in his review paper but at date the NEF is not commonly used
for atmospheric studies.

The longwave radiative code represents a very significant part of the computer time requirements of most GCMs and some important simplifications have to be performed. For instance, LW radiative codes are often called at lower frequencies than the rest of the physics, although some key variables (humidity, clouds, etc.) may change significantly from one time step to the next. Another example: most LW radiative codes neglect the effects of scattering by cloud water droplets and/or large aerosols although some studies have suggested that this approximation may lead to non negligible errors (for instance \cite{Edwards01}-\cite{Dufresne02}). Even if scattering can be considered by some codes, this possibility is commonly turned off in order to save CPU time \cite{Lacis.Mishchenko:1995}.

The purpose of this paper is to show that the NEF may be a good framework to get some useful insight into radiative transfer within the atmosphere, particularly for identification of the radiative exchanges that contributes the most to atmospheric heating rates. In the long term, on the basis of such analysis, our final goal is to develop a NEF parameterization of the effects of scattering in the LW. We will make a short presentation of the formalism and of the Monte-Carlo Method that we have developed (\para{sec1}). The choice of a Monte-Carlo Method was motivated by its high flexibility. We will present some analysis for clear sky atmospheres (\para{sec2}) and for cloudy atmospheres (\para{sec3}) through which we further illustrate and quantify some classical results. Finally, we will analyze how scattering may affect radiative exchanges in cloudy atmospheres (\para{sec4}).\section{Net Exchange Formulation and Monte-Carlo Method} \label{para:sec1}

Most numerical methods for solving longwave atmospheric radiative transfer are based on the integro-differential form of the radiative transfer equation. Using the Net Exchange Formulation (NEF) based on Net Exchange Rates (NER) as first proposed by Green (1967) \cite{Green} may represent a very interesting alternative. In terms of physical images, the NEF allows a complete decomposition of net radiative budgets as sums of NERs between each atmospheric layer and ground, space, and the rest of the atmosphere. An application similar to the present one, (using a NEF toward radiative transfer parameterization in a GCM) was the subject of a previous work concerning the Martian atmosphere \cite{Dufresne03}. Cooling to space is known to be the dominant term of the cooling rate and the "cool to space" approximation has first been proposed by Rodgers and Walshaw \cite{Rodgers:66}. Dividing the cooling rate into a "cooling to space" term and an "exchange" term allows to build fast and precise algorithms \cite{Fels:75,Schwarzkopf:91}. Here we go further and we gather the cooling rate in individual exchange terms between each pair of layer of the atmosphere.

The Net Exchange Rate between two arbitrary elements $i$ and $j$ (either volume or surface elements) is defined as the energy rate emitted from element $i$ and absorbed by $j$, minus energy rate emitted from $j$ and absorbed by $i$. In the very general case of an inhomogeneous absorbing and scattering medium, NER have been formulated in a previous article \cite{eymet01}. In the case of a 1-dimensional atmospheric radiative transfer calculation (see \fig{atmos}(a)), expressing the monochromatic NER $\Psi_{ij,\nu}$ between two arbitrary atmospheric layers $i$ and $j$ is straight forward, as illustrated below:

\begin{figure}[htbp]
\centering
\mbox{
\subfigure[]{\begin{turn}{-90}\epsfig{figure=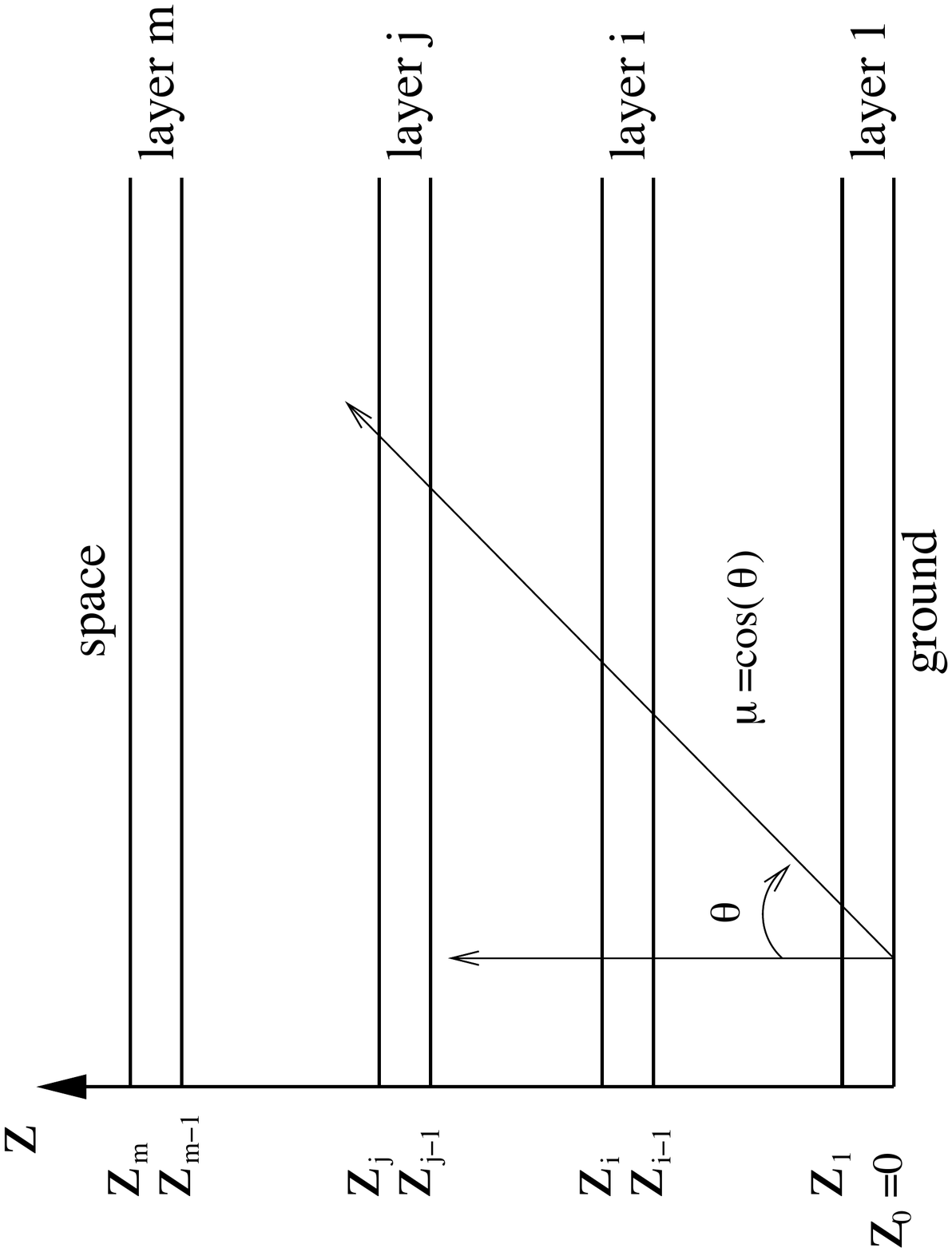,width=0.40\textwidth}\end{turn}}\quad
\subfigure[]{\begin{turn}{-90}\epsfig{figure=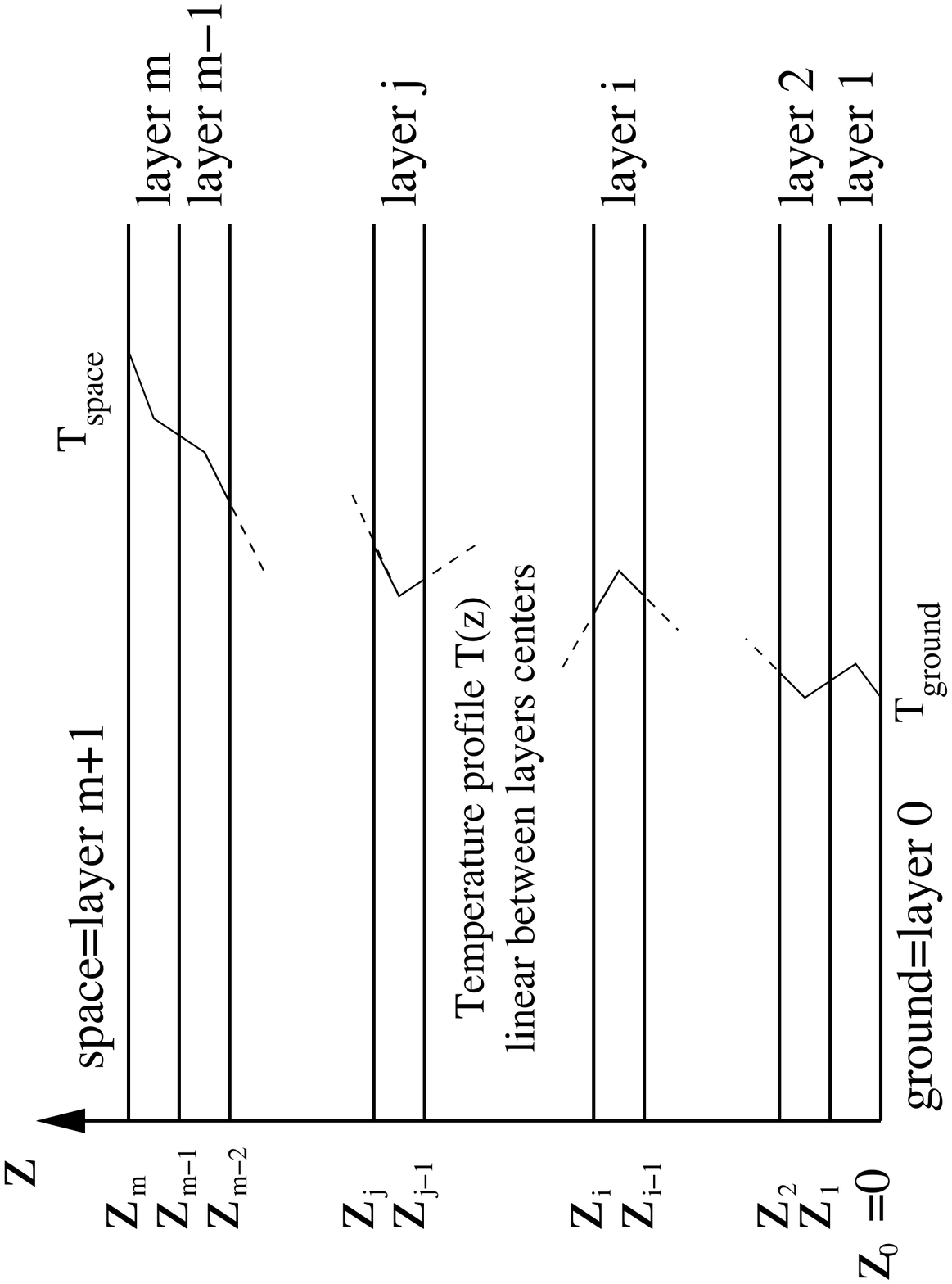,width=0.40\textwidth}\end{turn}}
}
\caption{(a): classical discretization of 1-D inhomogeneous atmosphere into $m$ homogeneous atmospheric layers~; (b): atmospheric configuration used for radiative transfer computations.}
\label{fig:atmos}
\end{figure}

\begin{equation}
\Psi_{ij,\nu}=\int_{2\pi}d\Omega\int_{z_{i-1}}^{z_{i}}\frac{dz}{\mu}\int _{z_{j-1}}^{z_{j}}\frac{dz^{\prime}}{\mu} \mu k_{a,\nu}(z)k_{a,\nu}(z^{\prime})\Bigl(B_{\nu}(z)-B_{\nu}(z^{\prime})\Bigr)exp\Bigl(-\int_{z}^{z^{\prime}}\frac{k_{a,\nu}(z^{\prime \prime})}{\mu}dz^{\prime \prime}\Bigr)
\label{eq:psi_ij}
\end{equation}

As this expression makes clear, if layer $i$ is warmer than layer $j$, $\Psi_{ij,\nu}$ is the rate at which layer $i$ is cooled by all radiative interactions with layer $j$ (and reciprocally, the rate at which $j$ is warmed by interactions with $i$).

Similarly, the monochromatic NER $\Psi_{ig,\nu}$ and $\Psi_{is,\nu}$ respectively between layer $i$ and the ground, and between layer $i$ and space, are given below for purely absorbing atmospheres:

\begin{equation}
\Psi_{ig,\nu}=\int_{2\pi}d\Omega\int_{z_{i-1}}^{z_{i}}\frac{dz}{\mu}\mu k_{a,\nu}(z)\Bigl(B_{\nu}(z)-B_{\nu}(ground)\Bigr)exp\Bigl(-\int_{0}^{z} \frac{k_{a,\nu}(z^{\prime})}{\mu}dz^{\prime}\Bigr)
\label{eq:psi_ig}
\end{equation}

\begin{equation}
\Psi_{is,\nu}=\int_{2\pi}d\Omega\int_{z_{i-1}}^{z_{i}}\frac{dz}{\mu}\mu k_{a,\nu}(z)\Bigl(B_{\nu}(z)-B_{\nu}(space)\Bigr)exp\Bigl(-\int_{z}^{H}\frac{k_{a,\nu}(z^{\prime})}{\mu}dz^{\prime}\Bigr)
\label{eq:psi_is}
\end{equation}

$z_{i}$ represents the top altitude of layer $i$, $k_{a,\nu}(z)$ is the monochromatic absorption coefficient at altitude $z$, and $B_{\nu}(z)$ is the monochromatic blackbody intensity at altitude $z$. Also, $d\Omega=sin(\theta)d\theta d\phi$, with $\theta$ and $\phi$ being the classical spherical coordinates. 

Still for illustration purposes, with the assumption of a uniform temperature profile within each atmospheric layer, \eq{psi_ij} reduces to:
\begin{equation}
\Psi_{ij,\nu}=\xi_{ij,\nu}\Bigl(B_{\nu}(i)-B_{\nu}(j)\Bigr)
\label{eq:xidB}
\end{equation}

where $B_{\nu}(i)$ and $B_{\nu}(j)$ are monochromatic blackbody intensities in layers $i$ and $j$. $\xi_{ij,\nu}$ is a monochromatic optico-geometric factor. With the assumption of a purely absorbing media, let $\tau_{i,j,\nu}$ be the monochromatic transmittivity between $z_{i}$ and $z_{j}$, $\xi_{ij,\nu}$ may then be expressed as a function of these monochromatic transmittivities:
\begin{equation}
\xi_{ij,\nu}=\tau_{i,j-1,\nu}-\tau_{i-1,j-1,\nu}-\tau_{i,j,\nu}+\tau_{i-1,j,\nu}
\label{eq:xi}
\end{equation}

Considering \eq{xi}, it is easy to see that $\xi_{ij,\nu}>0$~; thus, the sign of $\Psi_{ij,\nu}$ depends only on the sign of $B_{\nu}(i)-B_{\nu}(j)$ (still within the limits of the assumptions of a uniform temperature profile across each atmospheric layer and of a purely absorbing media).

During the last few years a series of articals \cite{Fournier04,Fournier03,Tesse01} have investigated use of the NEF within a Monte Carlo Framework.  NEF based Monte-Carlo algorithms have been improved in terms of sampling laws adjustment in order to bypass various numerical convergence difficulties encountered in quasi-isothermal configurations \cite{Fournier04}, purely absorbing optically thick media \cite{amaury02} and both absorbing and scattering optically thick media \cite{eymet01}.

For the present study, the choice of using a Monte-Carlo Method for computing NERs was mainly motivated by its high flexibility. In particular, the Monte-Carlo Method (MCM) is one of the numerical methods that can be easily adapted to perform NER computations \footnote{This is the case of all methods based on an integral form of the Radiative Transfer Equation, but differential methods are generally not adapted to NER evaluations}. Another main advantage of the MCM is its ability to compute, in addition to each radiative transfer quantity, a numerical uncertainty over this result (in fact, the statistical standard deviation), which makes the MCM a reference method. Finally, it has recently been shown \cite{amaury01} that sensitivities of any quantity to any parameter of the problem could be computed using the MCM with very low extra computing costs.

This paper is not intended to provide new theoretical developments concerning Monte-Carlo numerical algorithms. The results shown in the rest of this article have been obtained using the Monte-Carlo algorithm presented in \cite{eymet01}, for the particular case of the longwave radiative transfer in the Earth's atmosphere. Algorithmic details, and especially the methodological developments that allowed the development of an efficient algorithm to address Earth atmospheric radiation in the whole IR range (despite of the difficulties encountered by standard MCM at high optical thicknesses) can be found in the above references. The purpose of this paper is to present longwave radiative transfer computation results in some common terrestrial configurations, and to analyze these results with the emphasis on the effect of scattering by water and ice clouds.

The configuration is represented in \fig{atmos}(b). The inhomogeneous terrestrial atmosphere is divided into $m$ gaseous layers with homogeneous gas constituent concentrations, thus homogeneous optical properties (absorption coefficient $k_{a}$, scattering coefficient $k_{d}$, single scattering albedo $\omega$, phase function asymmetry parameter $g$). Ground is denoted layer $0$ and space is layer $m+1$. The temperature profile $T(z)$ is considered as linear between layers centers. Temperature of space is fixed at $0 K$ (no downward intensity from space in the longwave spectral domain).

Computations presented in the following sections have been performed using the atmospheric radiative transfer Monte-Carlo algorithm with $10^{6}$ statistical events per atmospheric layer, which ensures that every result has been computed with a statistical standard deviation lower than 0.1$\%$.

\section{Clear sky results}
\label{para:sec2}

The results presented in this section have been obtained for clear sky configurations.  Longwave atmospheric optical properties are based on detailed information from the HITRAN2000 molecular line database, which contains over a million spectral lines for 36 different molecules \cite{Rothman01}.  To minimize computational expense this information is converted into a set of correlated-k optical depths \cite{Lacis01} at a relatively coarse spectral resolution using the SBMOD model \cite{Yang01}. The first step in the process is to evaluate the line-by-line spectral profile within each spectral bin and for each layer in the atmosphere.  Second, the profiles are sampled at a high-spectral resolution, and sorted by absorption strength, thereby providing a smooth, monotonically increasing representation of the absorption strength distribution in the spectral band.  The resulting absorption-sorted profiles are then assigned into as many as 16 quadrature bins.  This process is applied to each of 50 homogeneous layers in the model atmosphere.  For the calculations performed here, a spectral resolution of 20 $cm^{-1}$ was chosen to cover the spectral range between $100$ and $2500 cm^{-1}$ ($4$ and $100 \mu m$).  Two standard \cite{McClatchey01} atmospheric profiles are used: a Mid-Latitude Summer (MLS) and a Sub-Arctic Winter (SAW) profile, with standard gaseous species concentrations.

\begin{figure}[htbp]
\centering
\mbox{
\subfigure[]{\begin{turn}{-90}\epsfig{figure=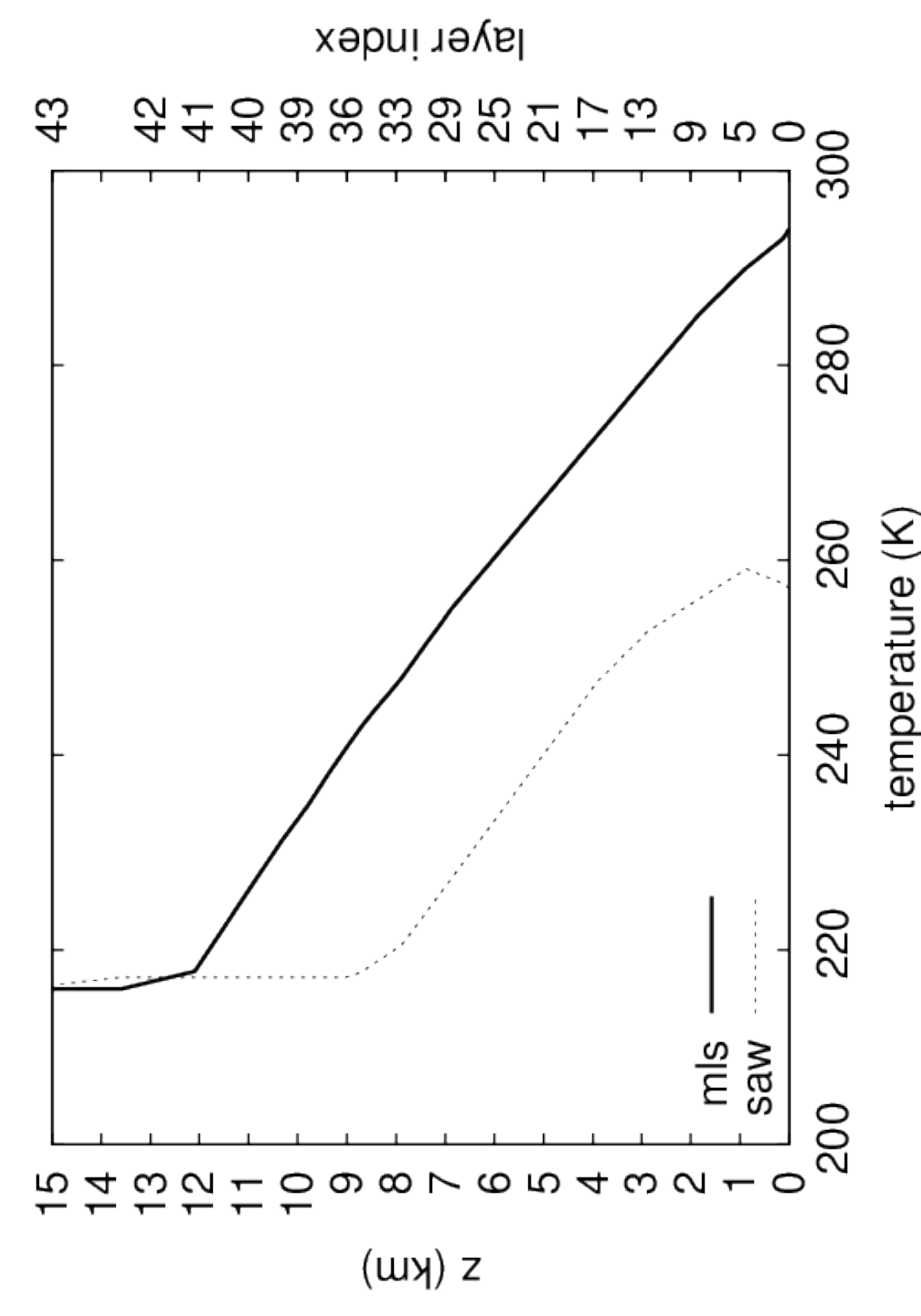,width=0.40\textwidth}\end{turn}}\quad 
\subfigure[]{\begin{turn}{-90}\epsfig{figure=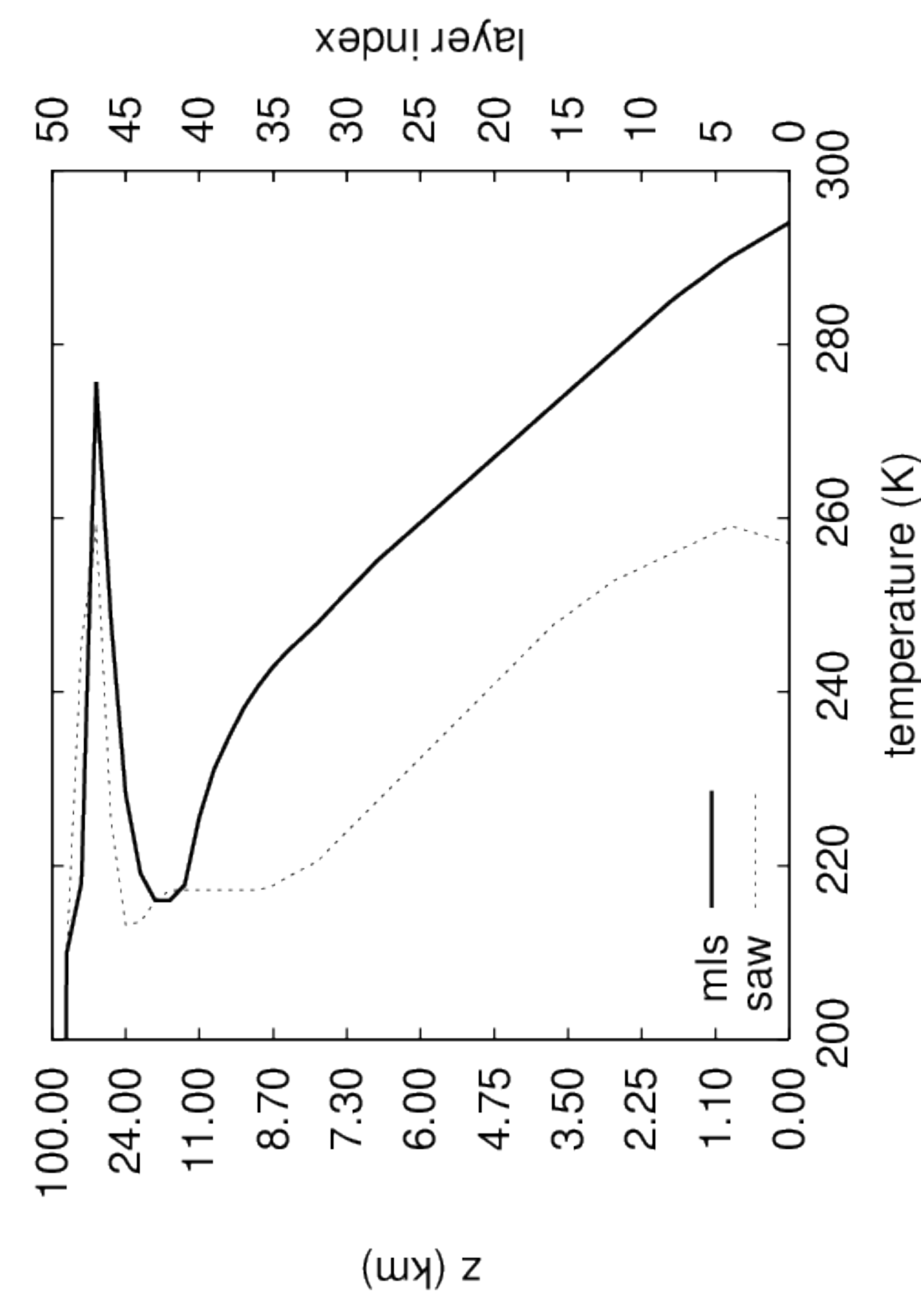,width=0.40\textwidth}\end{turn}}
}
    \caption{Temperature profiles (K) in a $M^{c}$ Clatchey Mid-Latitude Summer (MLS) atmosphere and in a Sub-Arctic Winter (SAW) atmosphere. Scale is (a) linear with altitude and (b) linear with layer index.}
\label{fig:temperature}
\end{figure}

\subsection{Heating Rates}

\begin{figure}[htbp]
\centering
\mbox{
\subfigure[MLS clear sky heating rate]{\begin{turn}{-90}\epsfig{figure=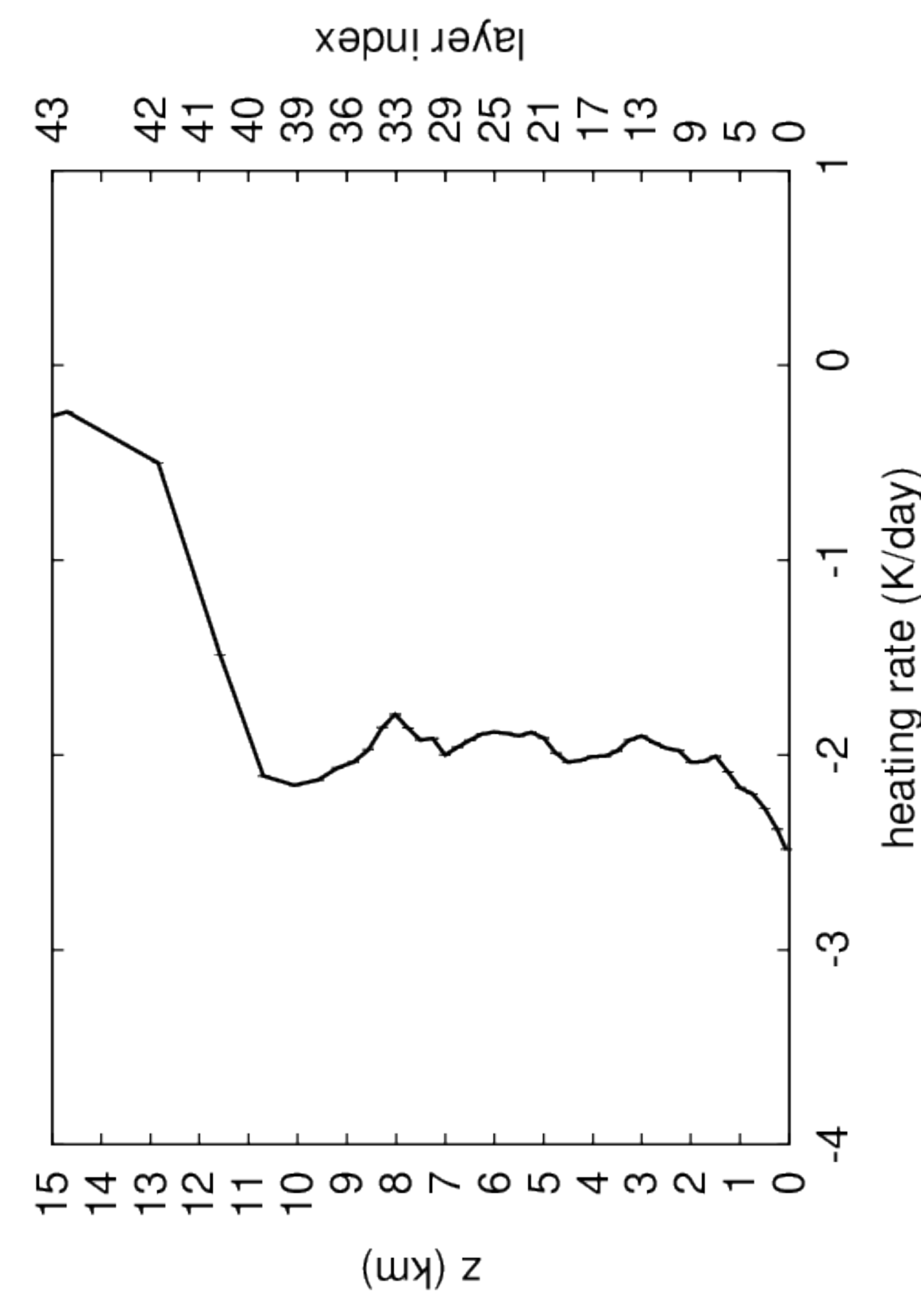,width=0.36\textwidth}\end{turn}}\quad
\subfigure[MLS clear sky NER matrix]{\begin{turn}{-90}\epsfig{figure=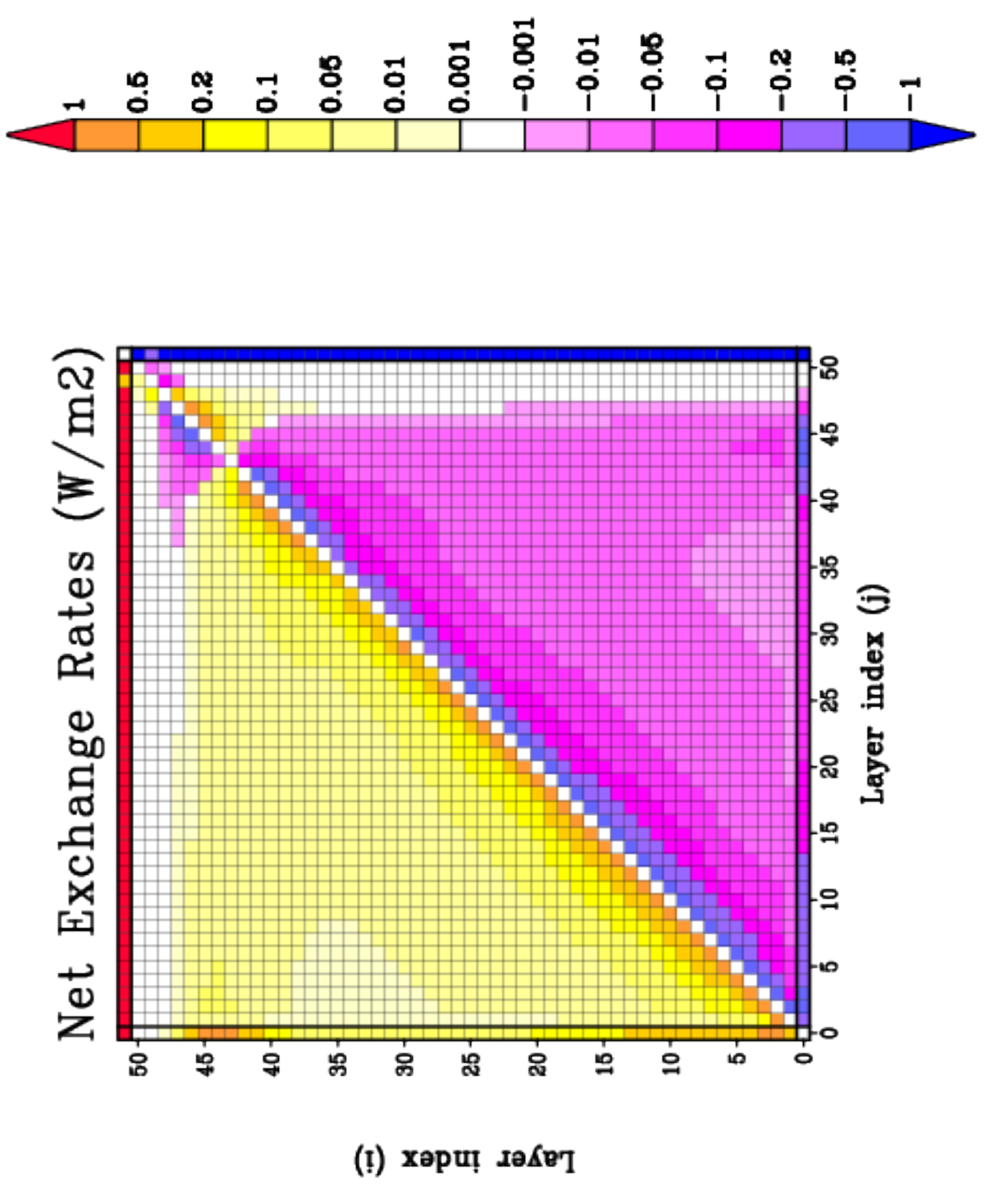,width=0.40\textwidth}\end{turn}}
}
    \caption{(a):  heating rate ($K/day$) for a clear sky MLS atmospheric profile, represented between ground and an altitude of $15 km$.(b): frequency integrated NERs ($W/m^{2}$) for the same atmospheric profile. The color of the square identified by layers numbers $i$ and $j$ represents the Net Exchange Rate between layers $i$ and $j$. Layer $0$ is the ground and layer $51$ represents space.}
\label{fig:mls_cs}
\end{figure}

\begin{figure}[htbp]
\centering
\mbox{
    \subfigure[SAW clear sky heating rate]{\begin{turn}{-90}\epsfig{figure=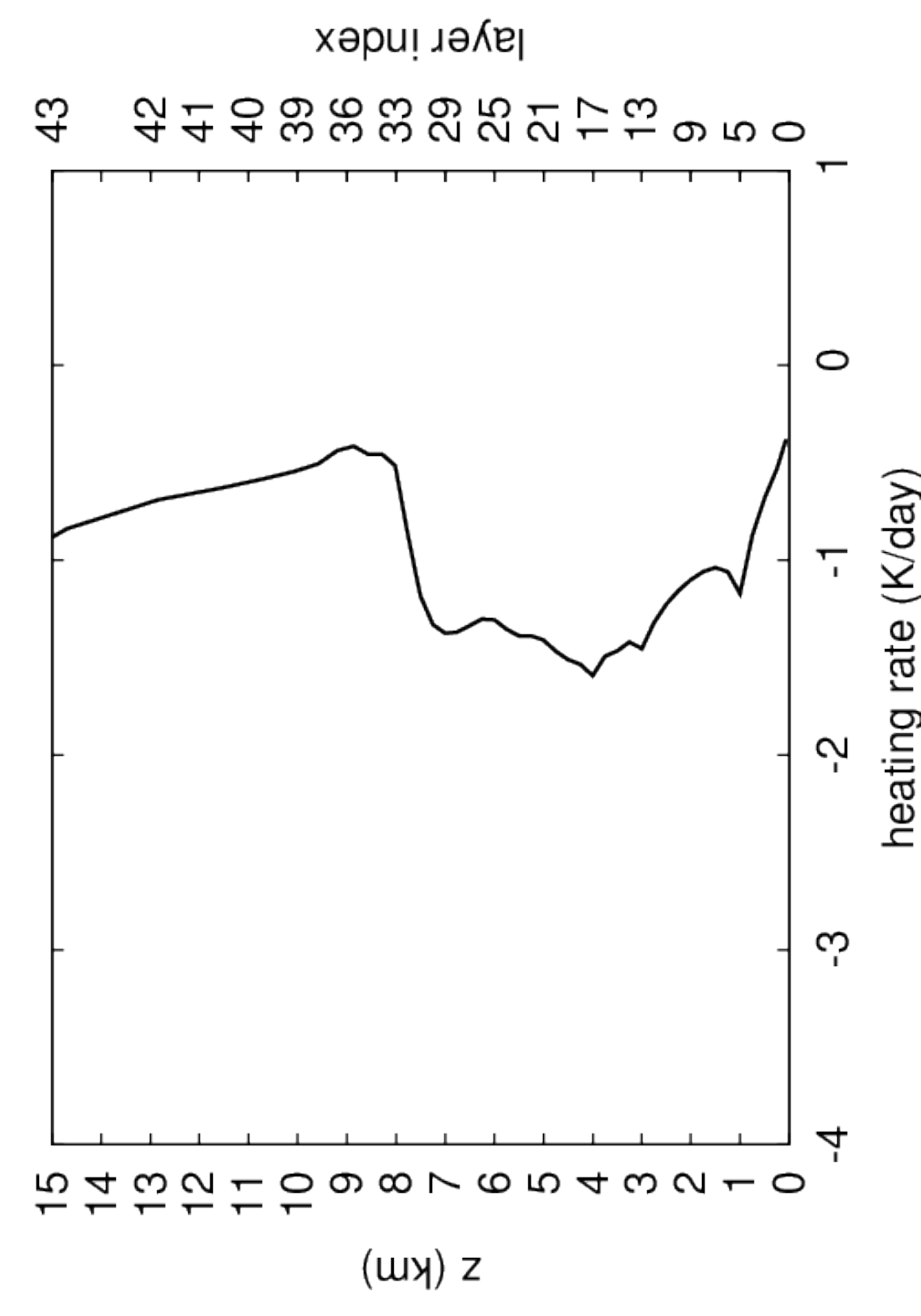,width=0.36\textwidth}\end{turn}}\quad
    \subfigure[SAW clear sky NER matrix]{\begin{turn}{-90}\epsfig{figure=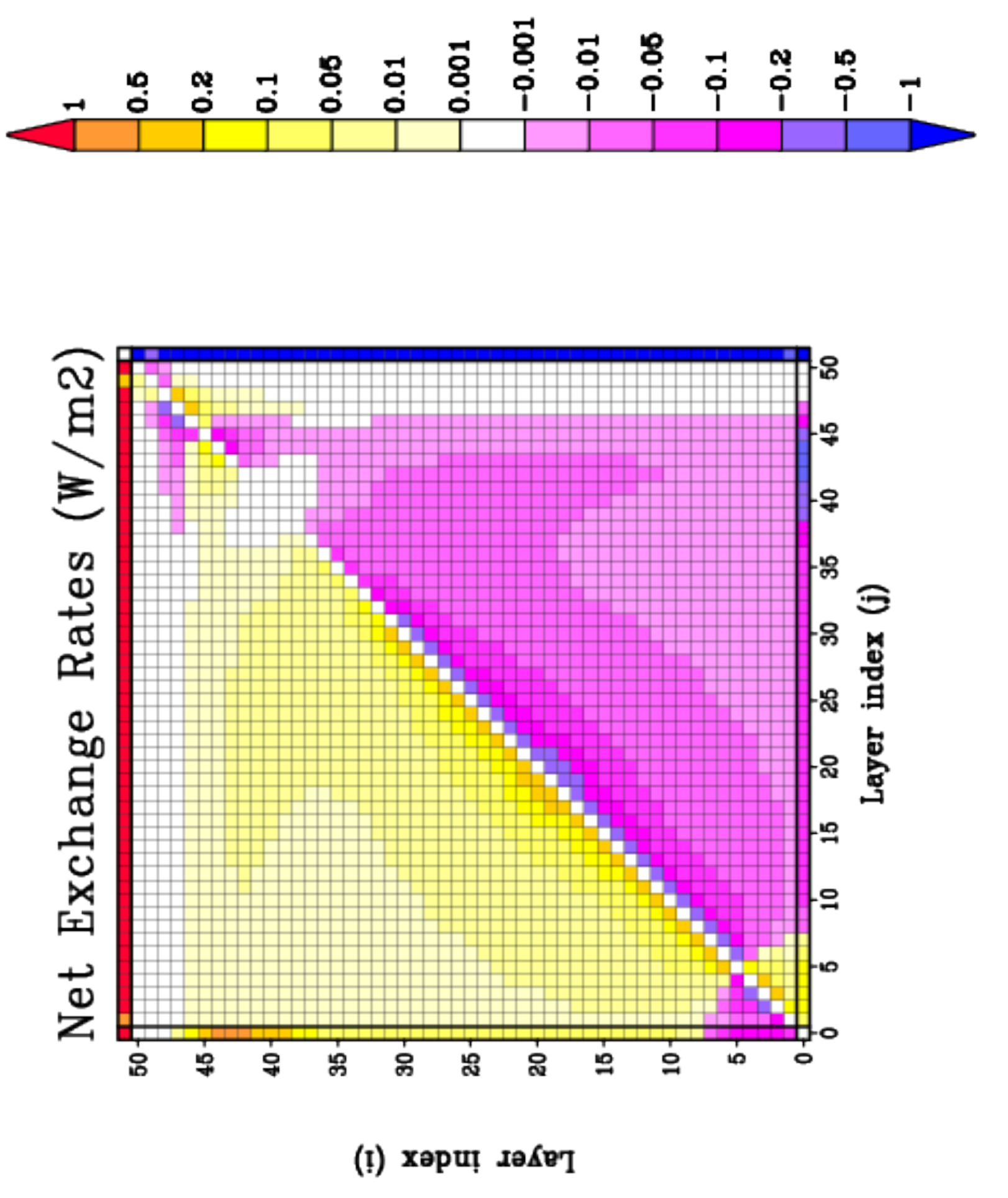,width=0.40\textwidth}\end{turn}}
}
    \caption{Same as \fig{mls_cs} for Sub-Arctic Winter atmospheric profile}
\label{fig:saw_cs}
\end{figure}

\fig{mls_cs}(a) and \fig{saw_cs}(a) show heating rates in $K/day$ integrated over the longwave spectral region for the MLS and SAW clear sky configurations. These results are comparable to results already published \cite{Fu01,Li01}. The heating rate in layer $i$, denoted $\chi_{i}$, may be interpreted as a sum of net exchanges, $\chi_{ij}$:
\begin{equation}
\chi_{ij}=\frac{g}{C_{p}}\frac{1}{\delta t}\frac{\Psi_{ij}}{\delta p_{i}} \label{eq:ki1}
\end{equation}

where $g$ is the gravity, $C_{p}$ is the heat capacity of air at constant pressure, $\delta t$ the length of the day, $\delta p_{i}$ the pressure difference between the top and the bottom of layer $i$, and $\Psi_{ij}$ the NER between layers $i$ and $j$. Within this framework, the total heating rate of layer $i$ is:
\begin{equation}
\chi_{i}=\sum_{j}\chi_{ij}
\label{eq:ki2}
\end{equation}

Of course, heating rates $\chi_{i}$ may also be computed from $\Psi_{i}=\sum_{j}\Psi_{ij}$ the total radiative budget of layer $i$:
\begin{equation}
\chi_{i}=\frac{g}{C_{p}}\frac{1}{\delta t}\frac{\Psi_{i}}{\delta p_{i}}
\label{eq:ki3}
\end{equation}

\begin{table}[ht] \centering
\myCaption{Top altitudes of atmospheric G.C.M. layers used in the vertical discretization}
\begin{tabular}{|c|c||c|c|} \hline
\textbf{Layer index} & \textbf{Top altitude (km)} & \textbf{Layer index} & \textbf{Top altitude (km)} \\ \hline
1 & 0.125 & 26 & 6.375 \\ \hline
2 & 0.375 & 27 & 6.625 \\ \hline
3 & 0.625 & 28 & 6.876 \\ \hline
4 & 0.875 & 29 & 7.126 \\ \hline
5 & 1.125 & 30 & 7.378 \\ \hline
6 & 1.375 & 31 & 7.632 \\ \hline
7 & 1.625 & 32 & 7.888 \\ \hline
8 & 1.875 & 33 & 8.149 \\ \hline
9 & 2.125 & 34 & 8.419 \\ \hline
10 & 2.375 & 35 & 8.704 \\ \hline
11 & 2.625 & 36 & 9.016 \\ \hline
12 & 2.875 & 37 & 9.371 \\ \hline
13 & 3.125 & 38 & 9.797 \\ \hline
14 & 3.375 & 39 & 10.339 \\ \hline
15 & 3.625 & 40 & 11.069 \\ \hline
16 & 3.875 & 41 & 12.094 \\ \hline
17 & 4.125 & 42 & 13.584 \\ \hline
18 & 4.375 & 43 & 15.795 \\ \hline
19 & 4.625 & 44 & 19.113 \\ \hline
20 & 4.875 & 45 & 24.114 \\ \hline
21 & 5.125 & 46 & 31.648 \\ \hline
22 & 5.375 & 47 & 42.964 \\ \hline
23 & 5.625 & 48 & 59.876 \\ \hline
24 & 5.875 & 49 & 85.000 \\ \hline
25 & 6.125 & 50 & 100.000 \\ \hline
\end{tabular}
\label{meshing}
\end{table}

\begin{table}[ht]
\centering
\myCaption{Frequency and wavenumber as function of narrow-band index}
\begin{tabular}{|c|c|c|} \hline
\textbf{Narrow-band index} & \textbf{frequency ($\mu m$)} & \textbf{wavenumber ($cm^{-1}$) } \\ \hline
1 & 4.000 & 2500 \\ \hline
15 & 4.504 & 2200 \\ \hline
26 & 5.000 & 2000 \\ \hline
35 & 5.494 & 1820 \\ \hline
43 & 6.024 & 1660 \\ \hline
49 & 6.493 & 1540 \\ \hline
55 & 7.042 & 1420 \\ \hline
64 & 8.064 & 1240 \\ \hline
76 & 10.000 & 1000 \\ \hline
86 & 12.500 & 800 \\ \hline
94 & 15.625 & 640 \\ \hline
101 & 20.000 & 500 \\ \hline
106 & 25.000 & 400 \\ \hline
110 & 31.250 & 320 \\ \hline
116 & 50.000 & 200 \\ \hline
118 & 62.500 & 160 \\ \hline
121 & 100.000 & 100 \\ \hline
\end{tabular}
\label{frequency}
\end{table}

The fact that individual heating rates may be interpreted as a sum of NER allows a complete decomposition of these heating rates~; in particular, this decomposition may, for instance, lead to the conclusion that the heating rate of a given layer $i$ is mainly driven by a small number of clearly identified NERs, and that the contribution of all others NERs is negligible.

\subsection{Net Exchange Rate matrices}

\fig{mls_cs}(b) and \fig{saw_cs}(b) show the Net Exchange Rates matrices (integrated over the longwave spectral region) in $W/m^{2}$. Each element of the matrix displays the intensity of the net exchange rate $\Psi_{ij}$ between each pair ($i$,$j$) of atmospheric layers. As an example of reading \fig{mls_cs}(b), consider layer $i=10$. The horizontal line of the matrix for layer $10$ shows the decomposition of the heating rate in layer $10$ in terms of NER contributions. Layer $10$ is for instance heated by the ground (layer $0$) and underlying warmer gas layers, and looses energy toward the colder layers above and toward space. Of course the matrix is antisymmetric because $\Psi_{ji}=-\Psi_{ij}$. Diagonal terms are null because there is no net exchange between a given layer and itself.

In \fig{mls_cs}(b), dominant terms are net exchanges between each atmospheric layer and the ground (first row and first column), each atmospheric layer and space (last row and last column), and also between close atmospheric layers (close to diagonal terms). Note an NER sign inversion in stratospheric layers: for instance, layer $45$ is heated by warmer layers $0$-$41$. However, layer 45 is also cooled from below by layers $42$-$44$, and heated from above by layers $46$-$48$. This is due to a MLS temperature profile inversion in the stratosphere.

Conclusions are similar for the SAW clear-sky configuration (\fig{saw_cs}(a)), except that two temperature inversions are noticeable: the first at $\approx 1 km$ altitude, an the second at $\approx 42 km$ altitude. These inversions in the temperature profile are visible in the NER matrix.  Also, a group of null near diagonal terms is visible for layers $38-43$. This is due to the fact that, in these layers, the temperature profile is a constant (see \fig{temperature}). 

\subsection{Spectral analysis}

The heating rate at a given altitude for a clear sky configuration is mainly due to heating from ground, cooling to space and exchange with adjacent atmospheric layers. From this conclusion, the total radiative budget $\Psi_{i,\nu}$ for each atmospheric layer $i$, in each narrowband centered around frequency $\nu$, has been decomposed into three main terms, the net exchange between atmospheric layer $i$ and ground $\Psi_{i,\nu}^{gas-ground}$, the net exchange between atmospheric layer $i$ and space $\Psi_{i,\nu}^{gas-space}$,and the net exchange between atmospheric layer $i$ and the rest of the atmosphere $\Psi_{i,\nu}^{gas-gas}$:
\begin{equation}
\Psi_{i,\nu}=\Psi_{i,\nu}^{gas-ground}+\Psi_{i,\nu}^{gas-space}+\Psi_{i,\nu}^{gas-gas}
\label{eq:psi_i_nu}
\end{equation} 

\begin{figure}[htbp]
\centering
\mbox{
    \subfigure[MLS clear sky $\Psi_{i,\nu}^{total}$]{\begin{turn}{-90}\epsfig{figure=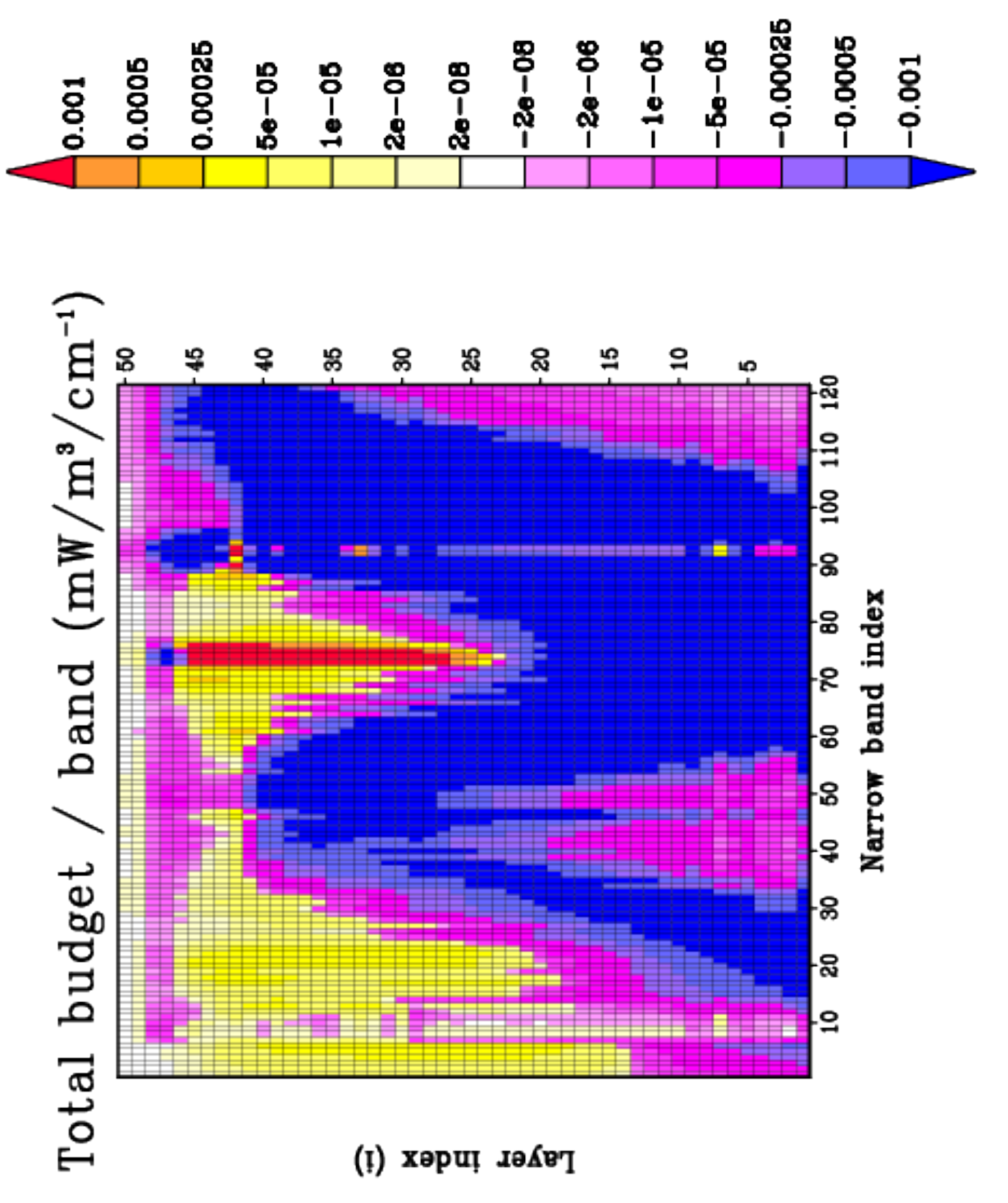,width=0.40\textwidth}\end{turn}}\quad
    \subfigure[MLS clear sky $\Psi_{i,\nu}^{gas-ground}$]{\begin{turn}{-90}\epsfig{figure=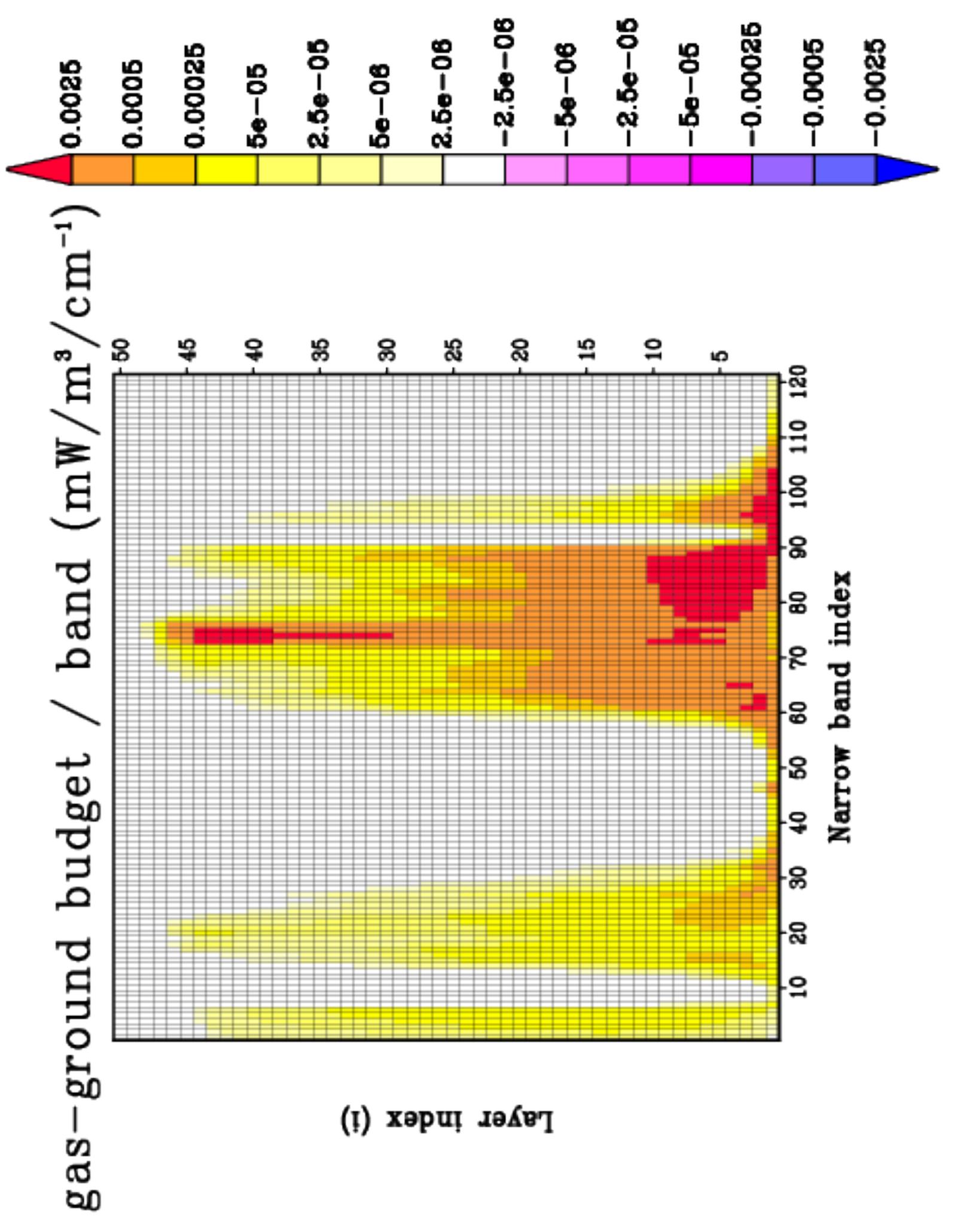,width=0.40\textwidth}\end{turn}}
}
\mbox{
    \subfigure[MLS clear sky $\Psi_{i,\nu}^{gas-space}$]{\begin{turn}{-90}\epsfig{figure=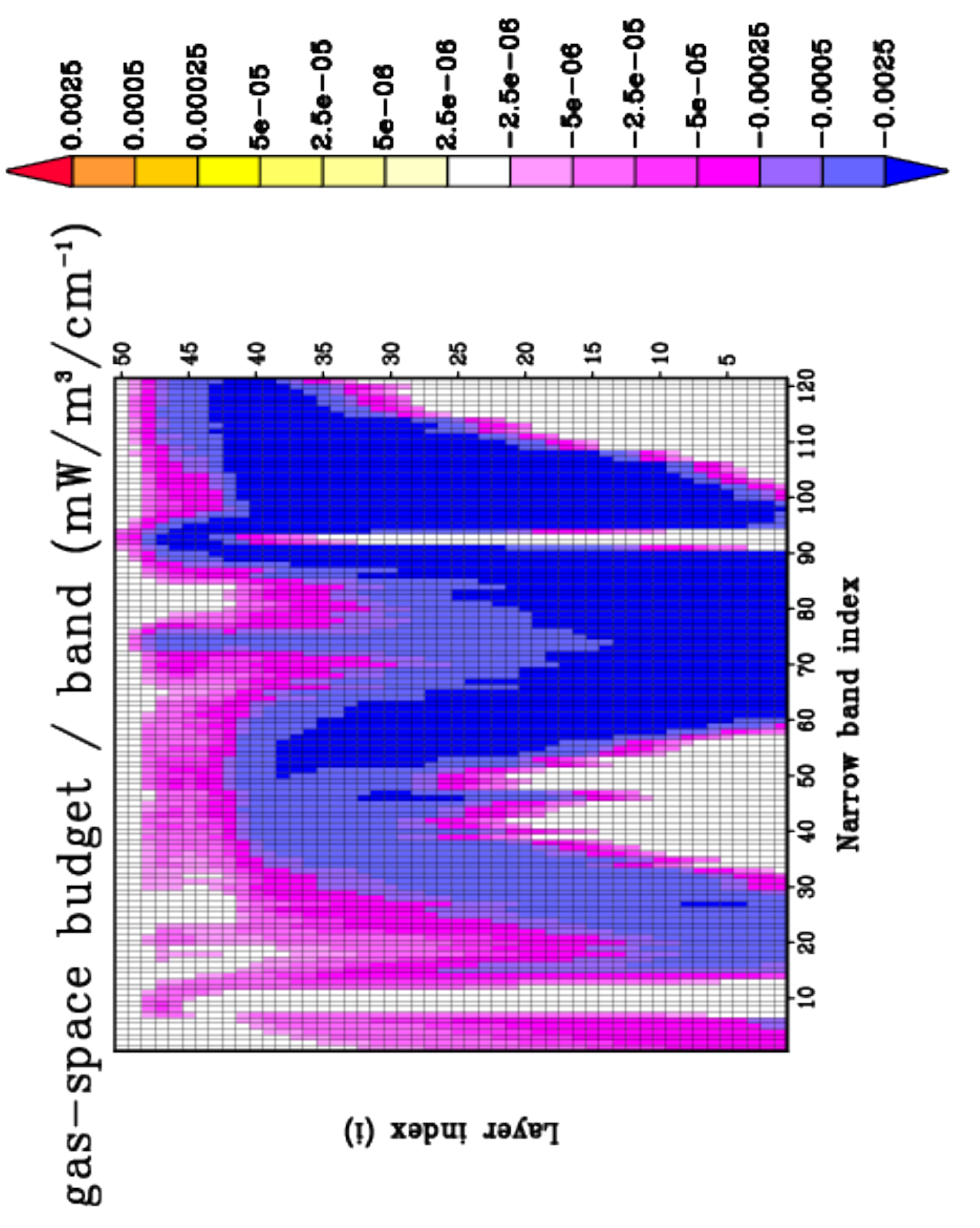,width=0.40\textwidth}\end{turn}}\quad
    \subfigure[MLS clear sky $\Psi_{i,\nu}^{gas-gas}$]{\begin{turn}{-90}\epsfig{figure=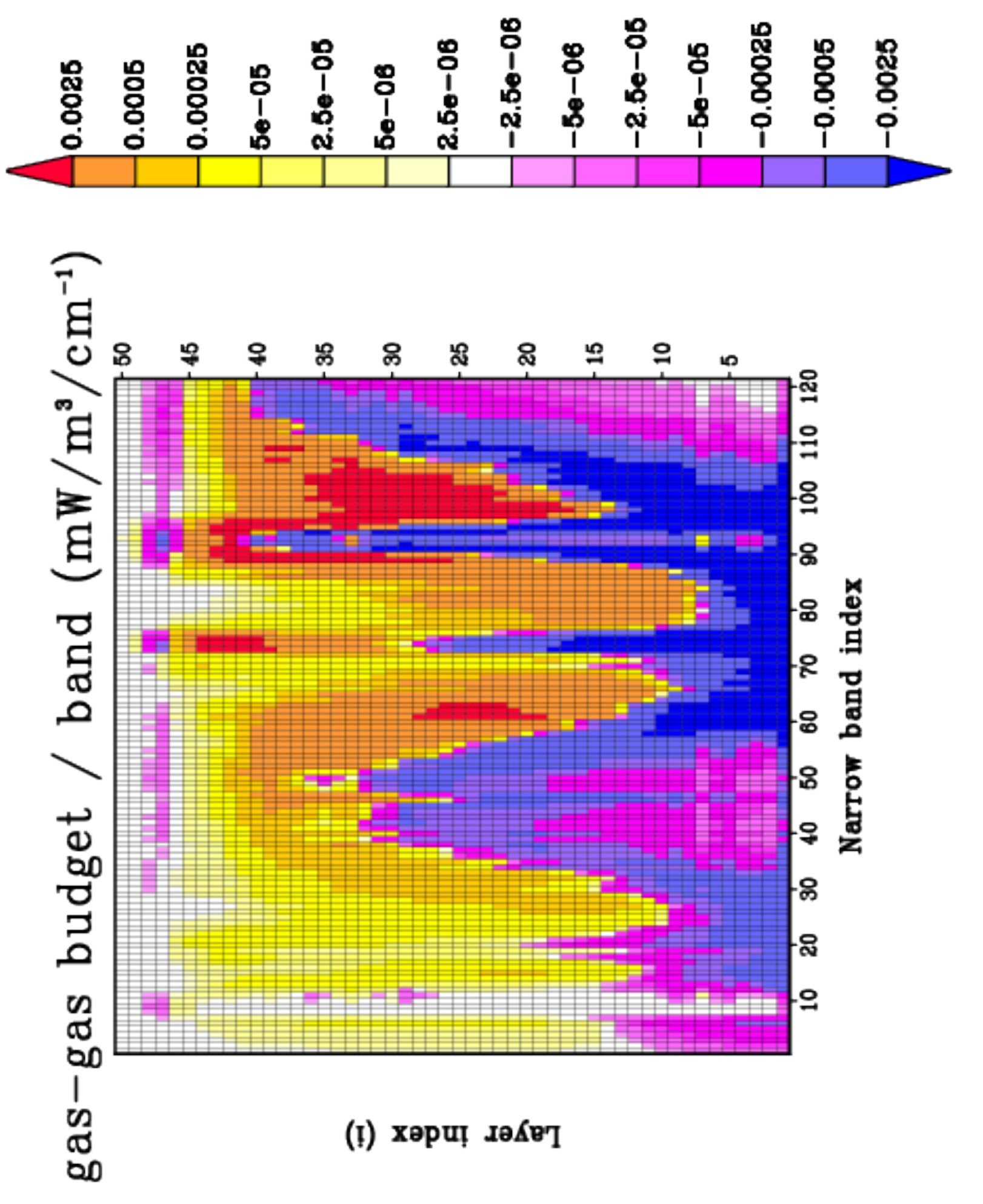,width=0.40\textwidth}\end{turn}}
}
    \caption{(a) Total radiative budget ($mW/m^{3}/cm^{-1}$) as a function of narrow-band index (see Table 2)  and atmospheric layer index, for a clear sky MLS configuration~; (b) Net exchange between each atmospheric layer and ground ($mW/m^{3}/cm^{-1}$)~; (c) Net exchange between each atmospheric layer and space ($mW/m^{3}/cm^{-1}$)~; (b) Net exchange between each atmospheric layer and the rest of the atmosphere ($mW/m^{3}/cm^{-1}$)}
\label{fig:bil_mls_cs}
\end{figure}

\begin{figure}[htbp]
\centering
\mbox{
    \subfigure[SAW clear sky $\Psi_{i,\nu}^{total}$]{\begin{turn}{-90}\epsfig{figure=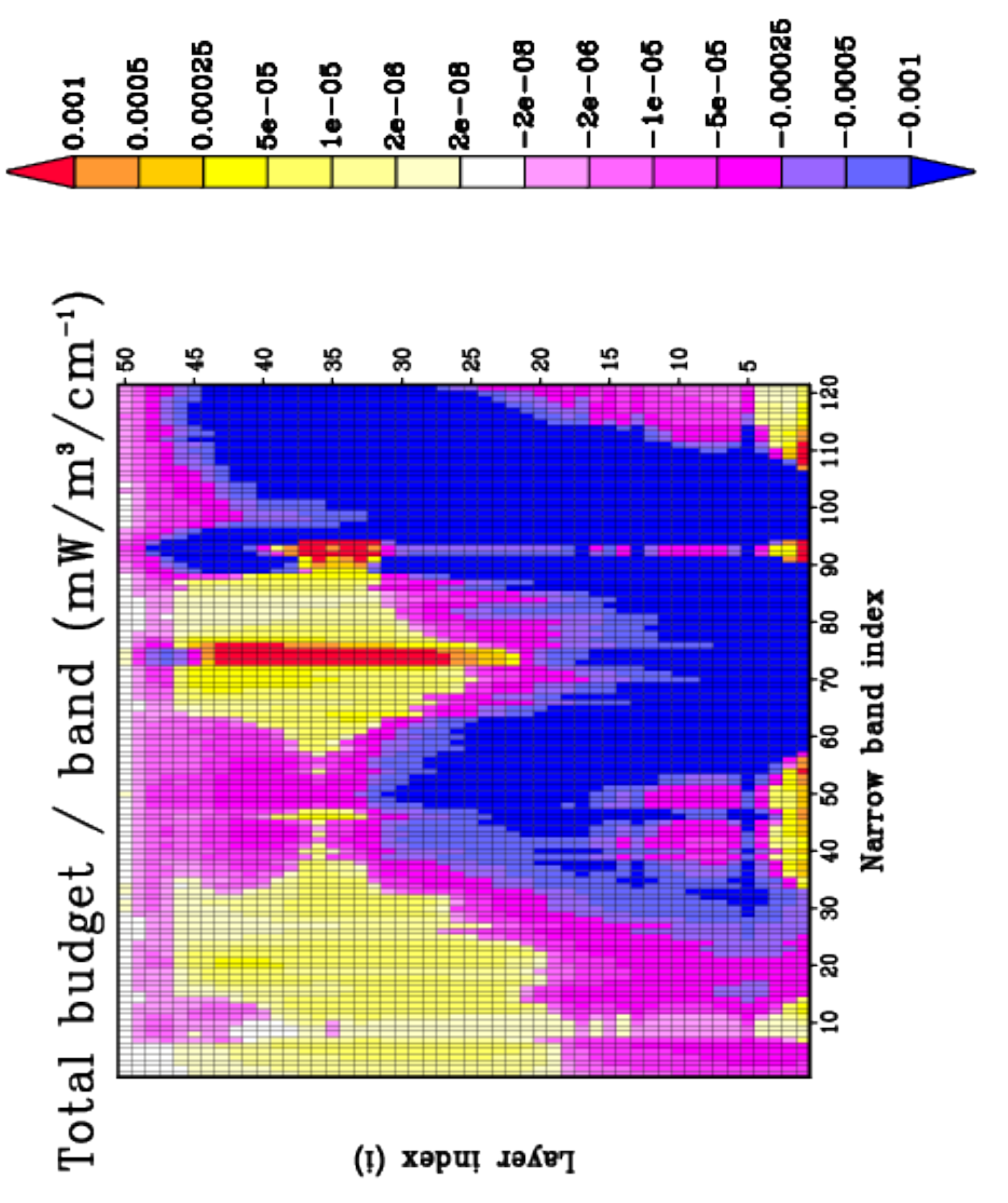,width=0.40\textwidth}\end{turn}}\quad
    \subfigure[SAW clear sky $\Psi_{i,\nu}^{gas-ground}$]{\begin{turn}{-90}\epsfig{figure=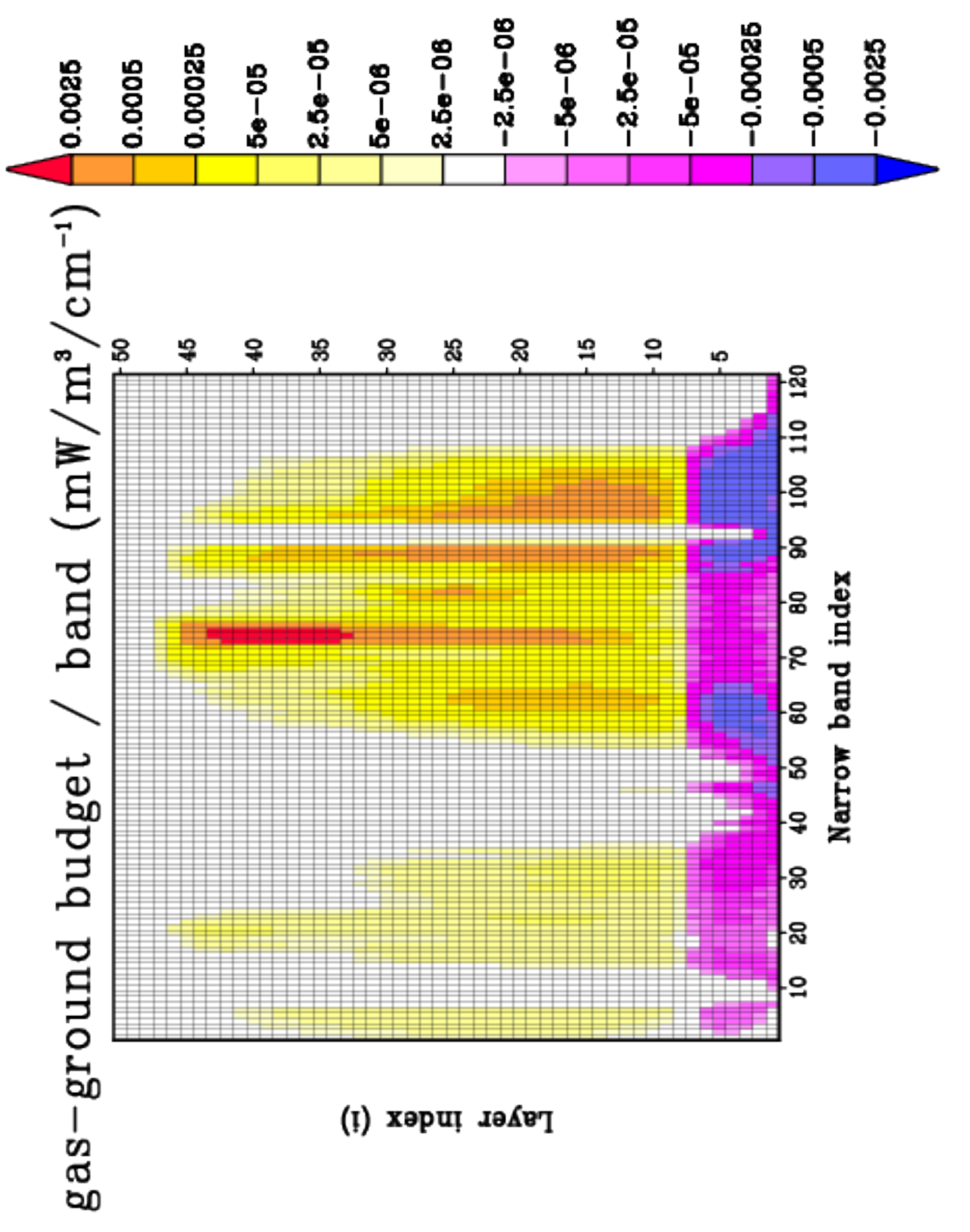,width=0.40\textwidth}\end{turn}}
}
\mbox{
    \subfigure[SAW clear sky $\Psi_{i,\nu}^{gas-space}$]{\begin{turn}{-90}\epsfig{figure=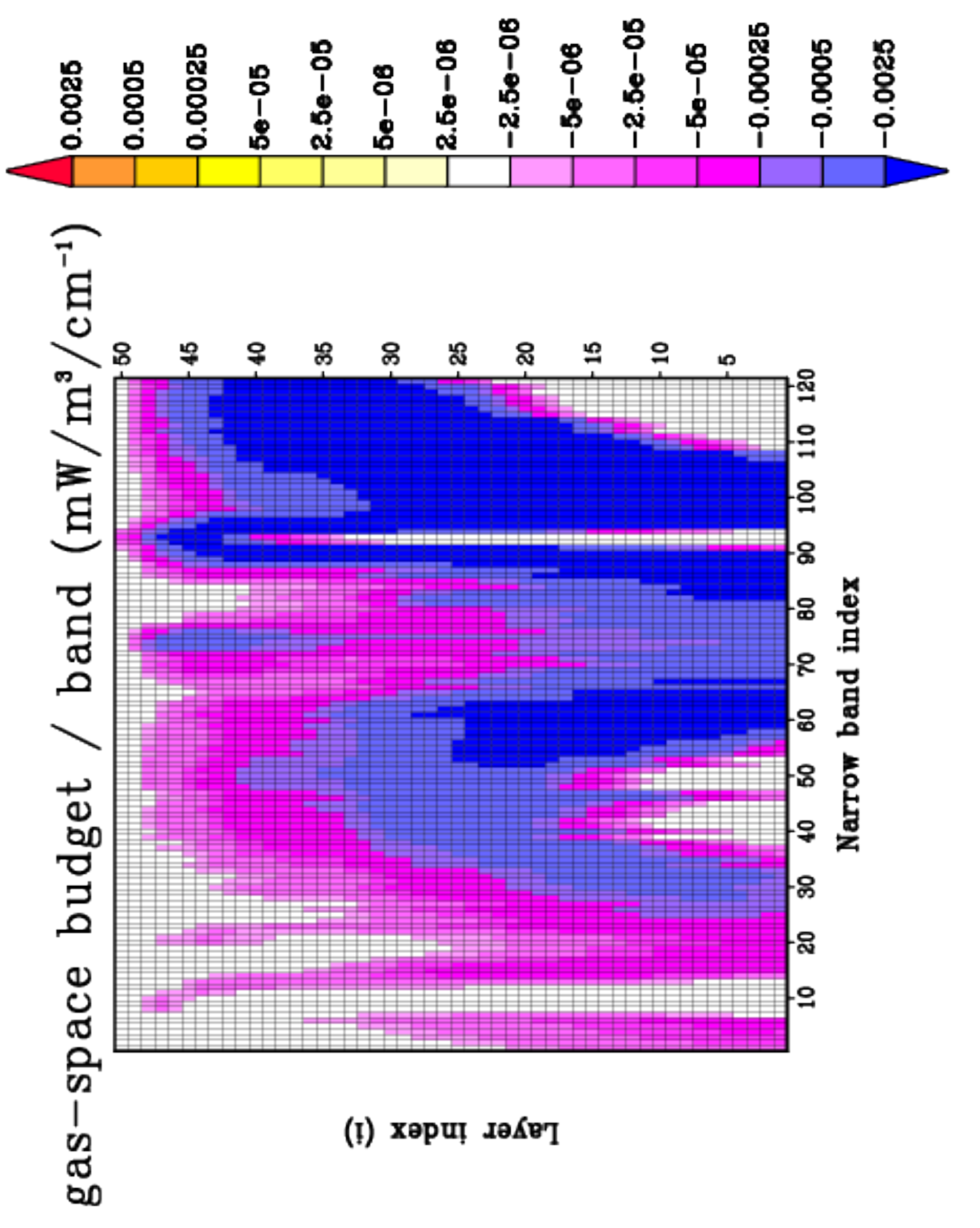,width=0.40\textwidth}\end{turn}}\quad
    \subfigure[SAW clear sky $\Psi_{i,\nu}^{gas-gas}$]{\begin{turn}{-90}\epsfig{figure=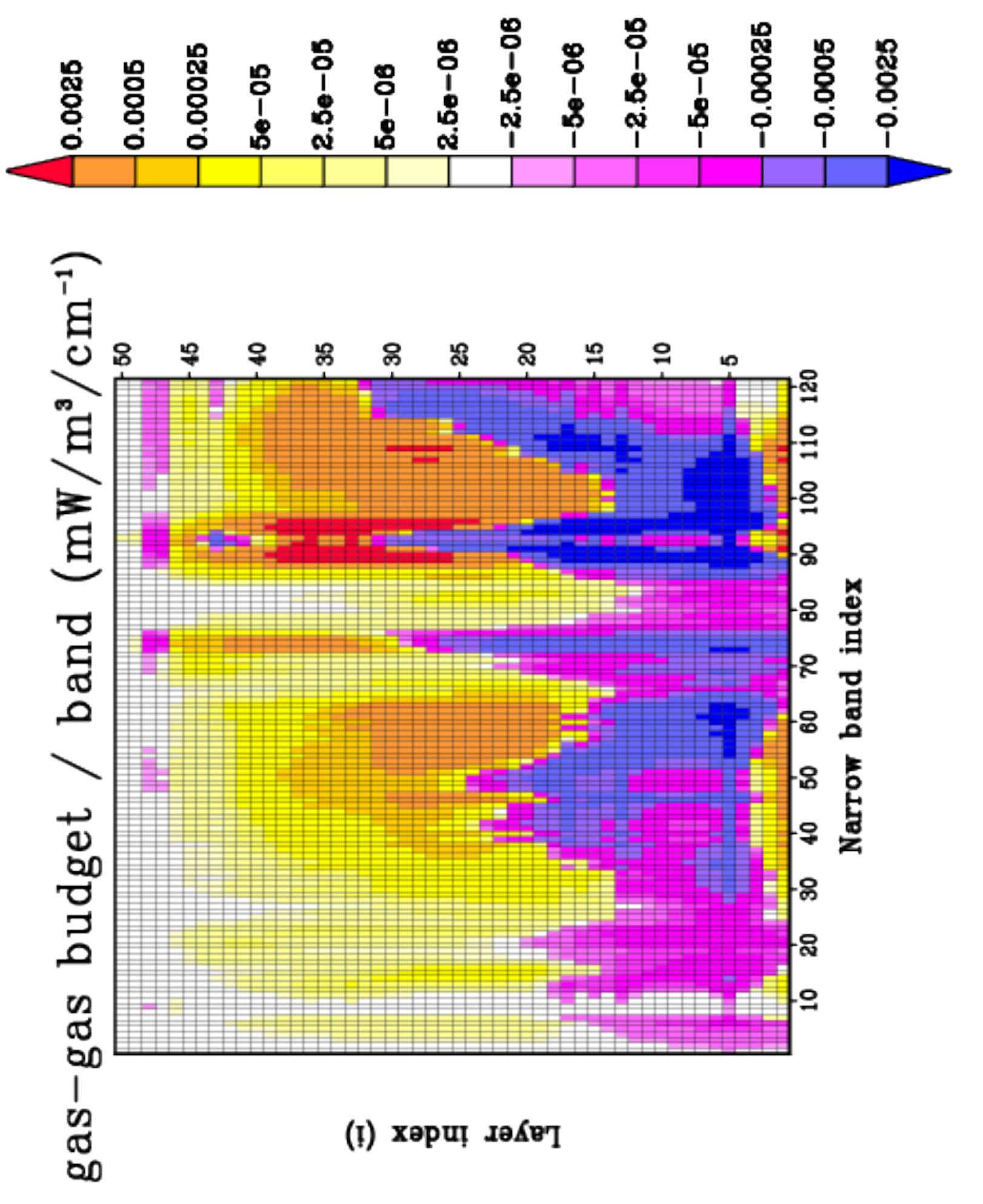,width=0.40\textwidth}\end{turn}}
}
    \caption{Same as \fig{bil_mls_cs} for a clear sky SAW configuration.}
\label{fig:bil_saw_cs}
\end{figure}

\fig{bil_mls_cs}(a) represents the total radiative budget as a function of altitude, for each spectral narrow-band of width 20 $cm^{-1}$ between $4$ and $100 \mu m$ (see table \ref{frequency} for correspondence between narrow-band index, frequency and wave-numbers) and may be compared with figures previously published by Clough and Iacono \cite{Clough:95}. \fig{bil_mls_cs}(b),\fig{bil_mls_cs}(c) and \fig{bil_mls_cs}(d) represent the different terms of the total radiative budget: net exchange between each atmospheric layer and ground, net exchange between each atmospheric layer and space, and net exchange between each atmospheric layer and the rest of the atmosphere. \fig{bil_saw_cs}(a)-\ref{fig:bil_saw_cs}(d) are identical to \fig{bil_mls_cs} for a clear sky SAW configuration. 

\fig{bil_mls_cs}(b) shows that the net exchange between each atmospheric layer and ground $\Psi_{i,\nu}^{gas-ground}$ is positive in a MLS configuration, because the ground  is heating all atmospheric layers: the ground temperature is higher than all gas temperatures. The heating effect is decreasing with altitude, because of absorption by lower atmospheric layers. The attenuation does not follow a strict exponential law, because of spectral correlations in the longwave range. Moreover, the heating from ground is dominant in low absorption spectral regions, such as the so-called ``atmospheric window'' which extends from $8$ to $13 \mu m$ at ground level (narrow-band index $63$-$82$), while net exchange between ground and atmospheric layers is null in strong absorption spectral regions, such as the $15 \mu m$ $CO_{2}$ band (narrow-band index $93$) and the water absorption regions (narrow-bands $30$ to $60$ and narrow-bands $100$ to $120$). In other absorption bands, such as the $9.2 \mu m$ ozone band (narrow-band $72$) and the $4.2 \mu m$ $CO_{2}$ band (narrow-band 10), $\Psi_{i,\nu}^{gas-ground}$ depends on the local variation of optical thickness with altitude. In a SAW configuration (\fig{bil_saw_cs}(b)), the temperature of gas is higher than the ground temperature in the low atmosphere (see \fig{temperature}). Thus, $\Psi_{i,\nu}^{gas-ground}$ is negative in the first $2$ km (first $7$ atmospheric layers)~; above $2$ km, the gas temperature drops below the ground temperature and $\Psi_{i,\nu}^{gas-ground}$ is positive.

The net exchange between each atmospheric layer and space $\Psi_{i,\nu}^{gas-space}$ (\fig{bil_mls_cs}(c) and \fig{bil_saw_cs}(c)) is negative, because space is cooler than all atmospheric layers, and $\Psi_{i,\nu}^{gas-space}$ is dominant in spectral  regions where variation of optical thickness with altitude is dominant. Net exchange between gas layers and space is only possible for layers that are ``visible'' from space. In low absorption spectral regions, all atmospheric layers can exchange radiation with space, while in strong absorption regions (such as water absorption bands), exchange with space is only possible for top layers.

The net exchange between each atmospheric layer and the rest of the atmosphere $\Psi_{i,\nu}^{gas-gas}$ (\fig{bil_mls_cs}(d) and \fig{bil_saw_cs}(d)) is positive for layers close to the ground, and negative for higher layers. The inversion altitude depends on the narrow-band, and is higher in strong absorption regions. $\Psi_{i,\nu}^{gas-gas}$ depends both on the local variation of optical thickness with altitude, and on the local variation of the temperature gradient with altitude. \section{Cloudy sky results} \label{para:sec3}

This section presents the results of computations held for cloudy configurations, with both MLS and SAW atmospheric profiles. For each atmospheric profile, four cloudy configurations have been examined (same configurations as in \cite{Fu01}): low cloud (water cloud extending from $1.0$ to $2.0 km$ for MLS, from $0.5$ to $1.5 km$ for SAW, with $LWC=0.22 g.m^{-3}$ and effective radius $r_{e}=5.89 \mu m$), middle cloud (water cloud extending from $4.0$ to $5.0 km$ for MLS, from $2.0$ to $3.0 km$ for SAW, with $LWC=0.28 g.m^{-3}$ and $r_{e}=6.20 \mu m$), high cloud (ice cloud extending from $10$ to $12 km$ for MLS, from $6$ to $8 km$ for SAW, with $IWC=0.0048 g.m^{-3}$ and effective diameter $D_{e}=41.5 \mu m$) and finally all three clouds simultaneously.

For these different configurations, this section will present NER matrices and heating rates, corresponding radiative budgets and their spectral decomposition.

\subsection{NER matrices and heating rates}

\begin{figure}[htbp]
\centering
\mbox{
    \subfigure[MLS middle-cloud heating rate]{\begin{turn}{-90}\epsfig{figure=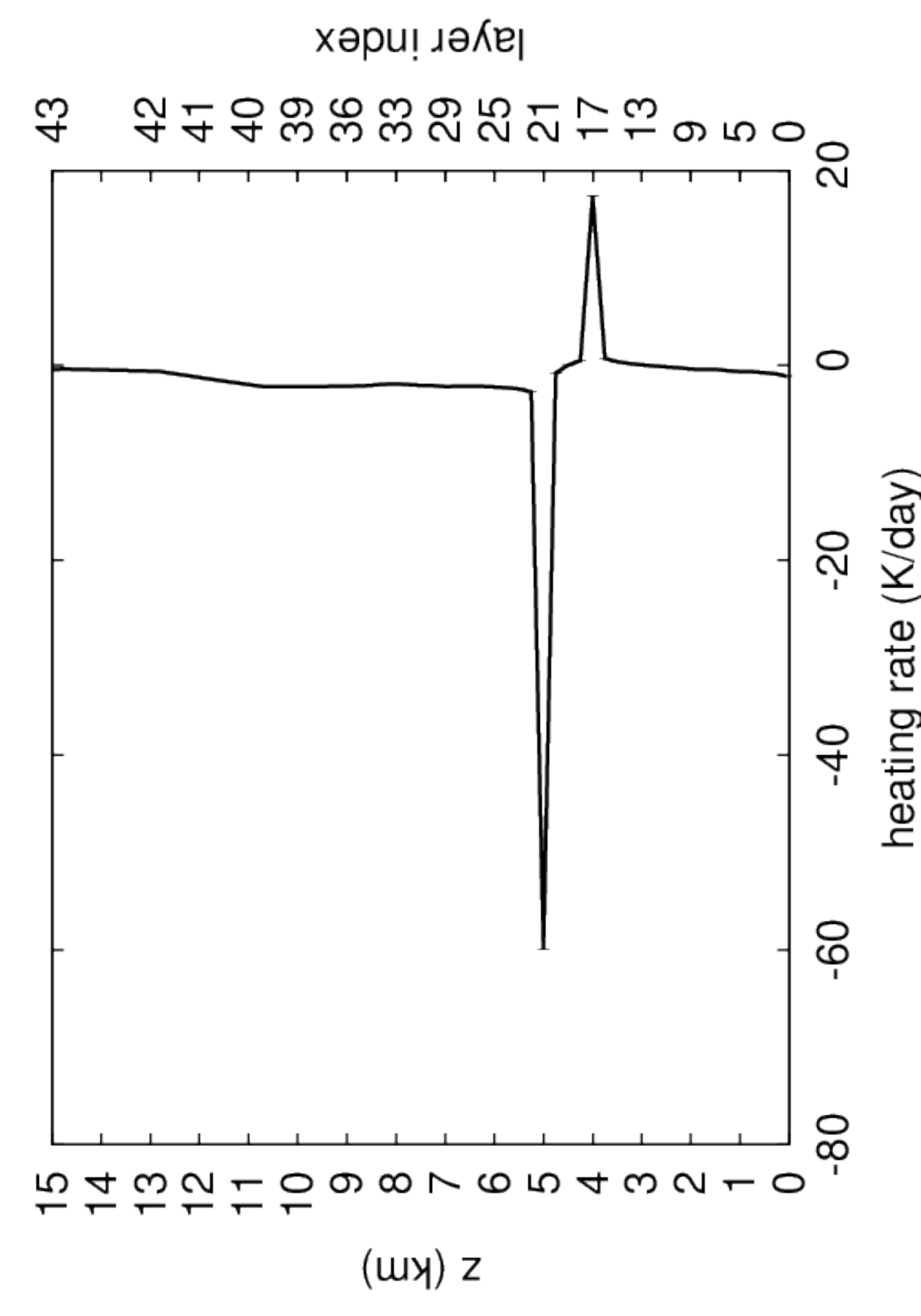,width=0.36\textwidth}\end{turn}}\quad
    \subfigure[SAW middle-cloud heating rate]{\begin{turn}{-90}\epsfig{figure=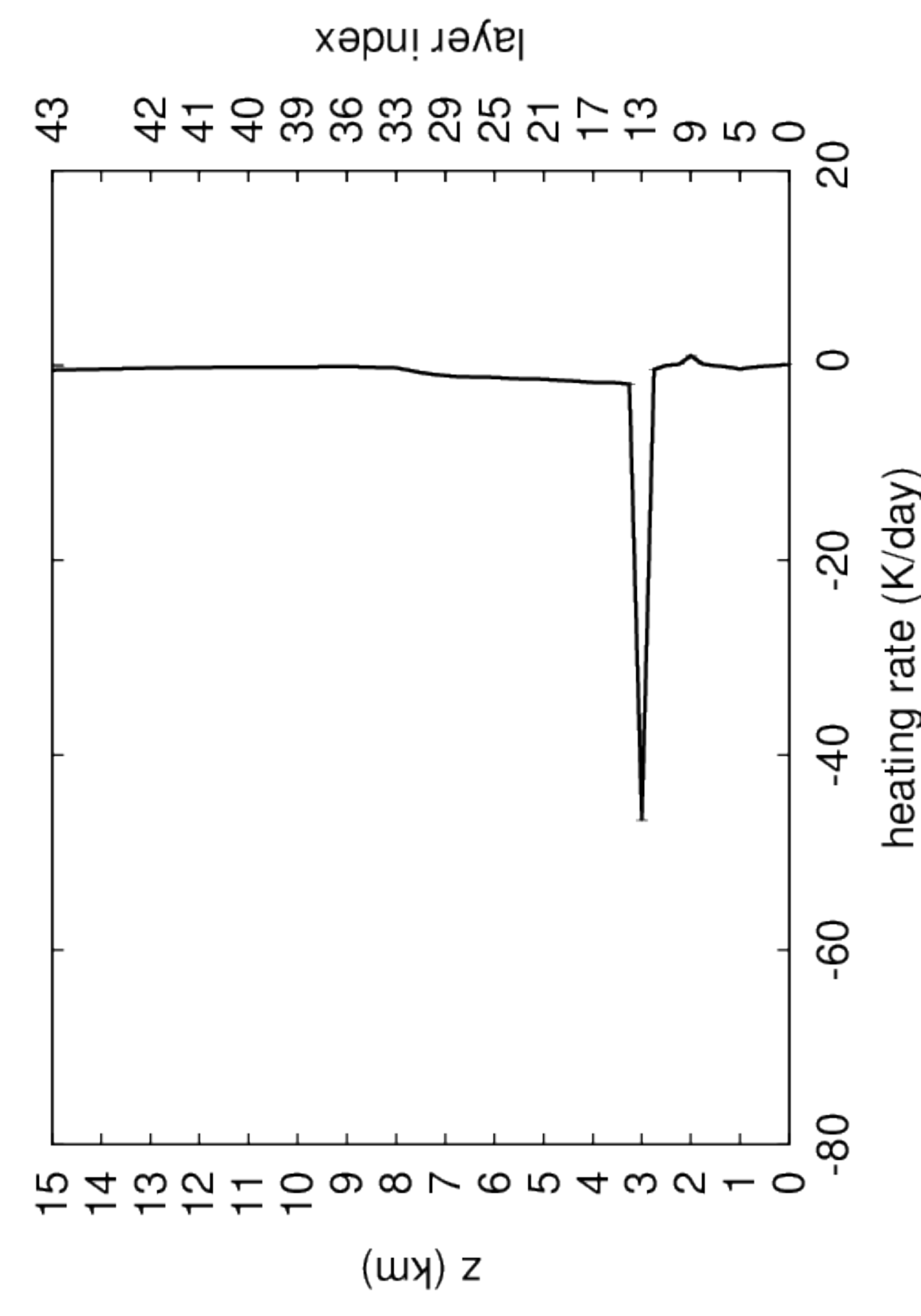,width=0.36\textwidth}\end{turn}}
}
\mbox{
    \subfigure[MLS middle-cloud NER matrix]{\begin{turn}{-90}\epsfig{figure=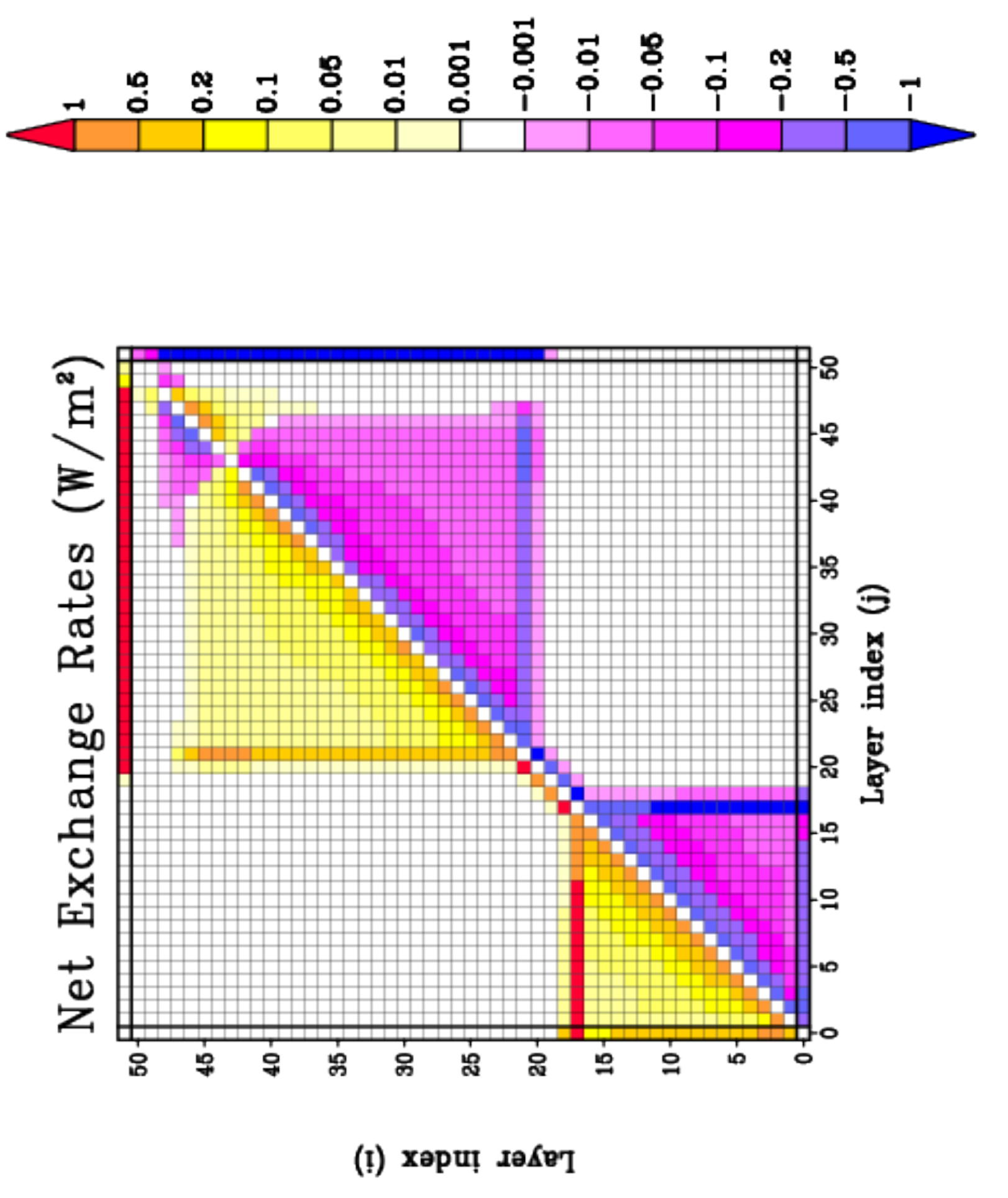,width=0.40\textwidth}\end{turn}}\quad
    \subfigure[SAW middle-cloud NER matrix]{\begin{turn}{-90}\epsfig{figure=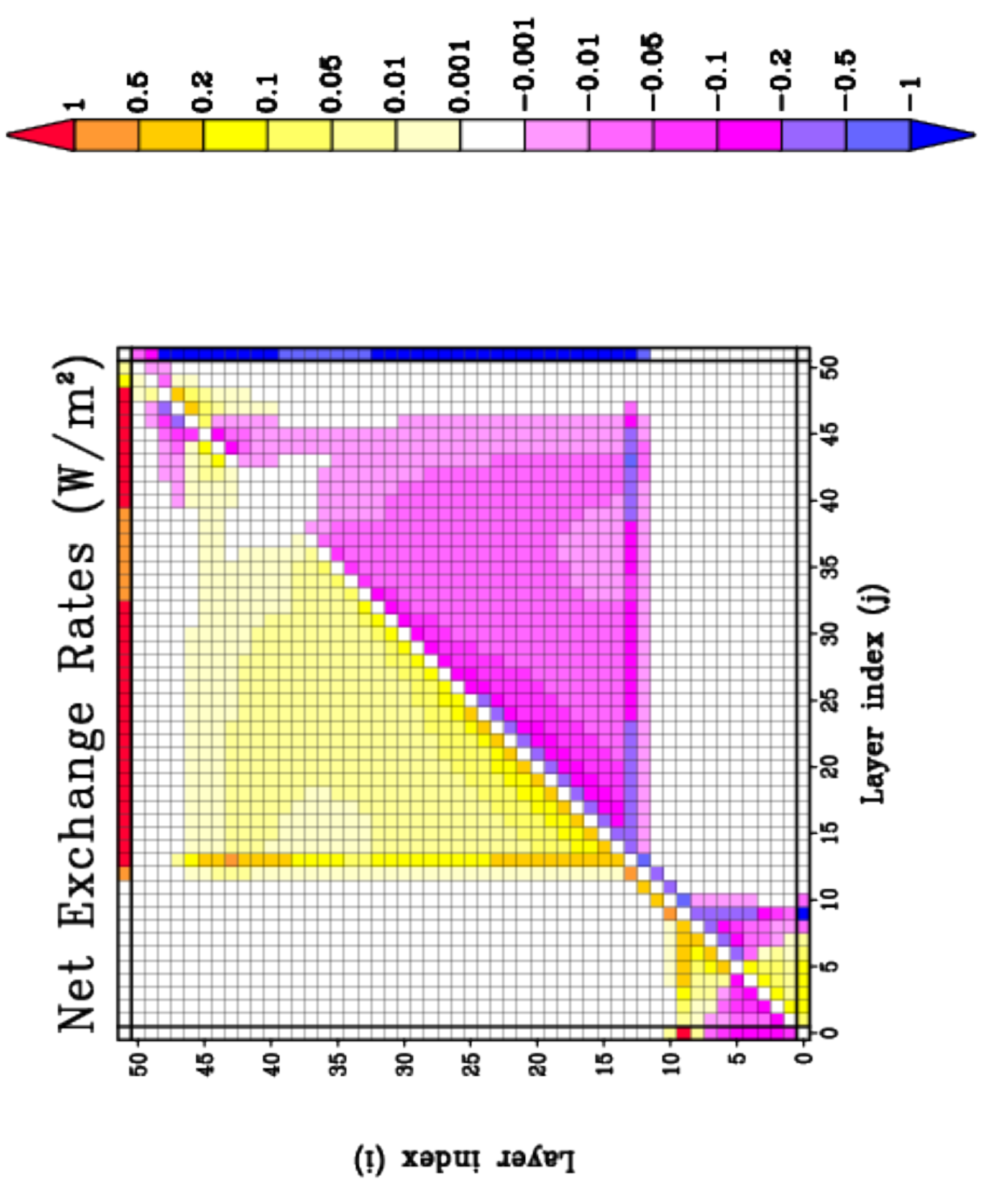,width=0.40\textwidth}\end{turn}}
}
    \caption{(a): heating rate ($K/day$) for MLS middle-cloud configuration (water cloud extending from $4.0$ to $5.0 km$ (layers $17-21$) with $LWC=0.28 g.m^{-3}$ and $r_{e}=6.20 \mu m$)~; (b): same as (a) for SAW atmosphere (water cloud extending from $2.0$ to $3.0 km$ (layers $8-13$) and same $LWC$ and $r_{e}$)~; (c): NER matrix $W/m^{2}$) in the middle-cloud configuration for a MLS atmospheric profile ~; (d): same as (c) for a SAW atmospheric profile.}
\label{fig:mc_ner}
\end{figure}

\begin{figure}[htbp]
\centering
\mbox{
    \subfigure[MLS high-cloud heating rate]{\begin{turn}{-90}\epsfig{figure=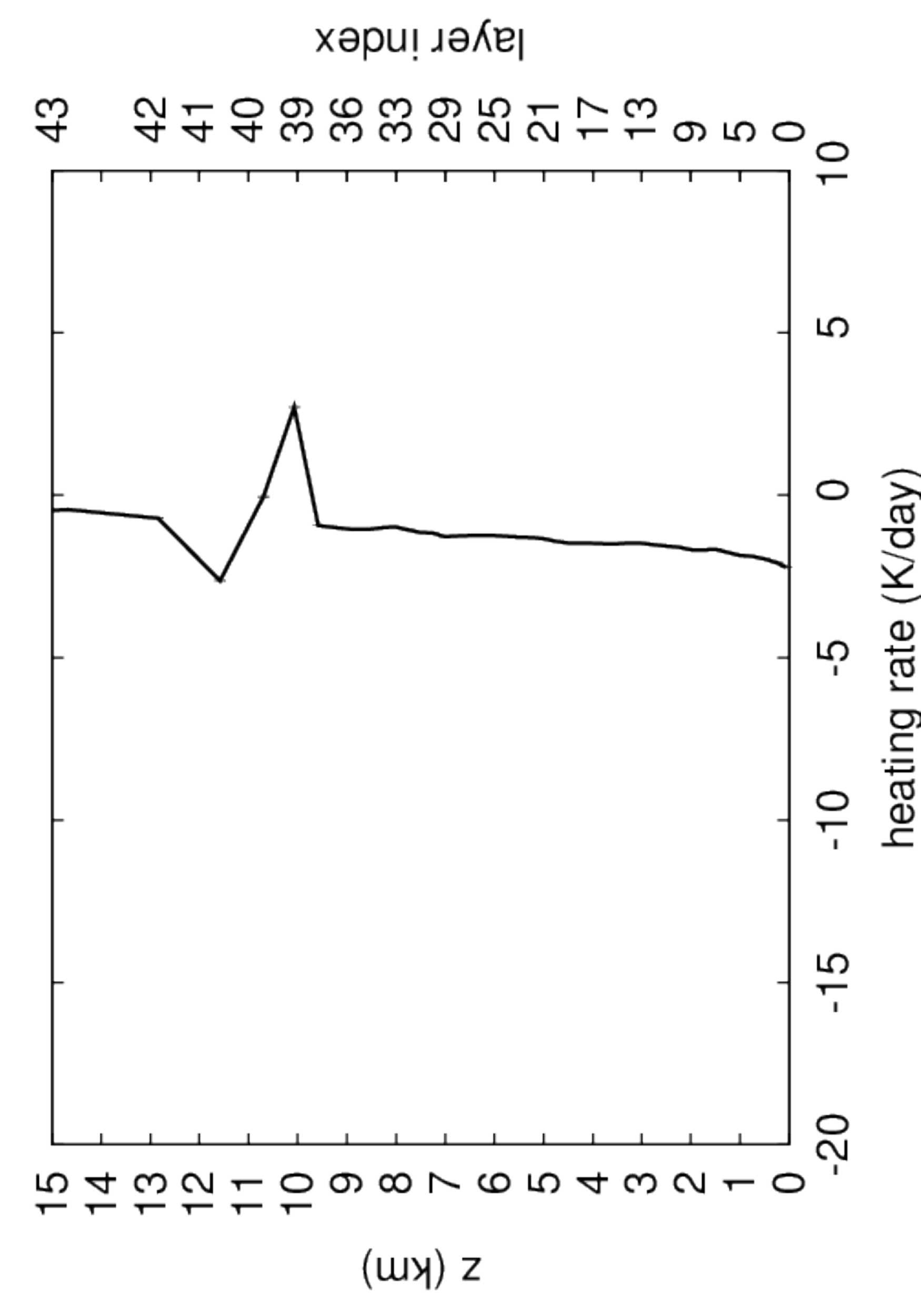,width=0.36\textwidth}\end{turn}}\quad
    \subfigure[SAW high-cloud heating rate]{\begin{turn}{-90}\epsfig{figure=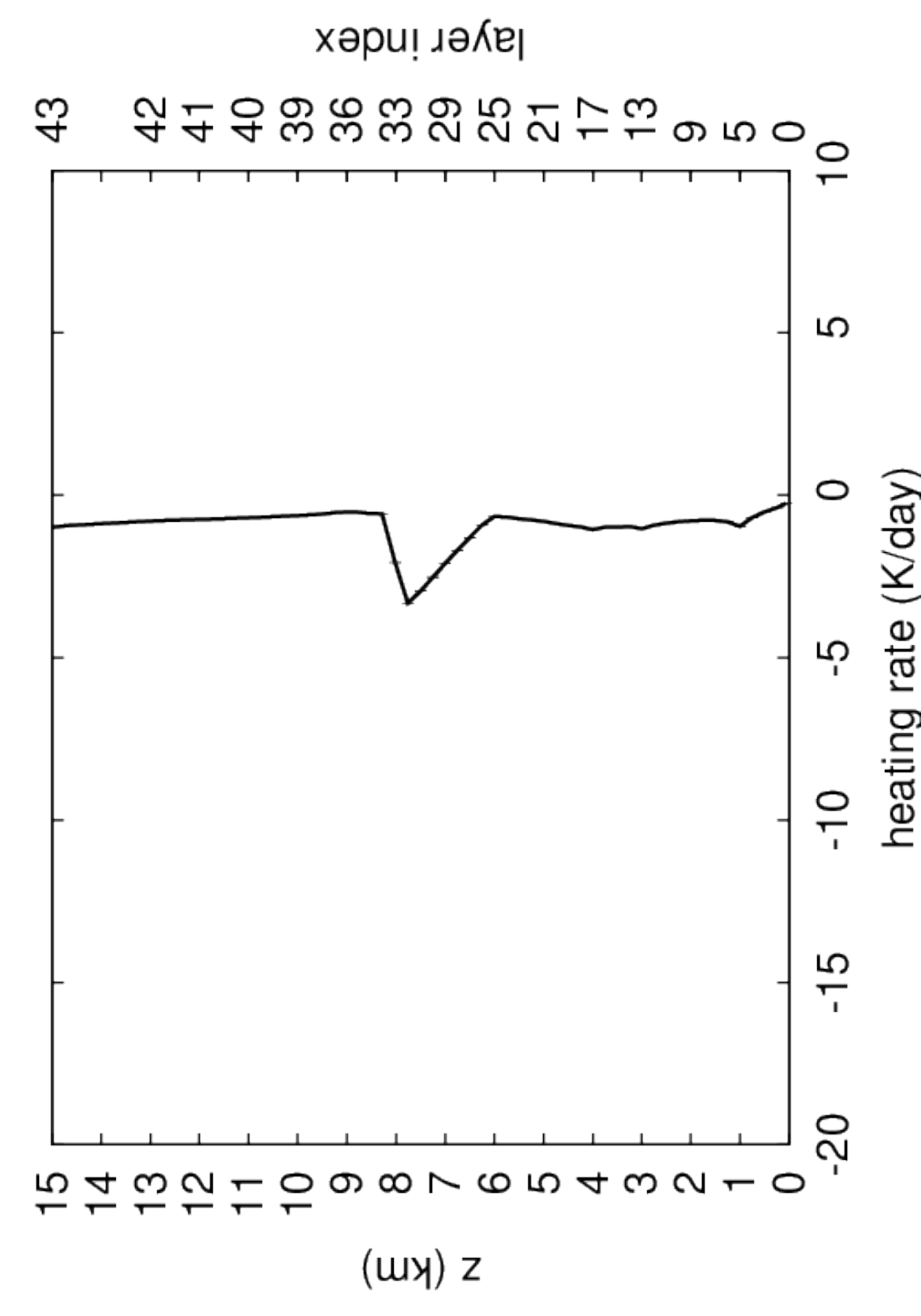,width=0.36\textwidth}\end{turn}}
}
\mbox{
    \subfigure[MLS high-cloud NER matrix]{\begin{turn}{-90}\epsfig{figure=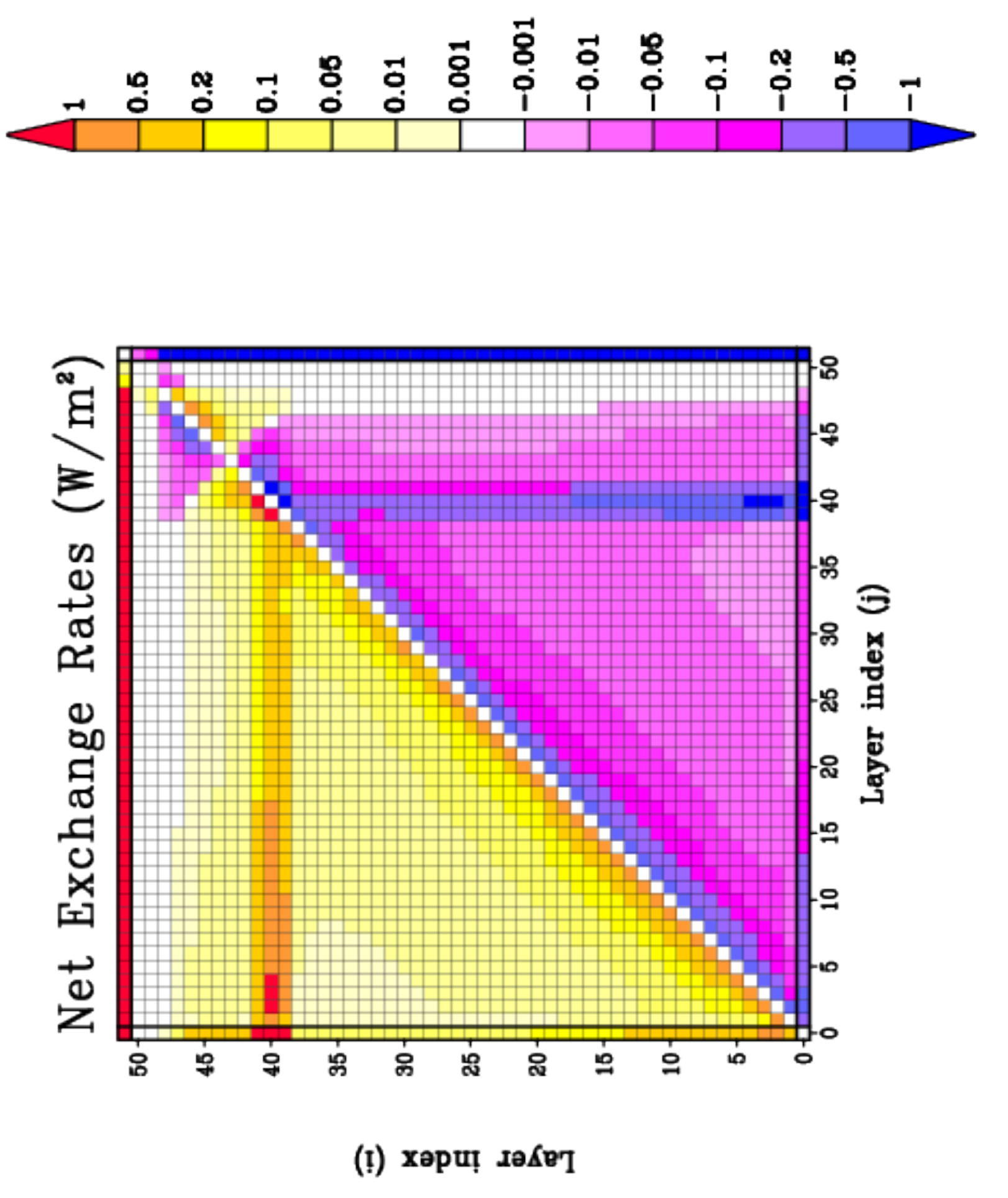,width=0.40\textwidth}\end{turn}}\quad
    \subfigure[SAW high-cloud NER matrix]{\begin{turn}{-90}\epsfig{figure=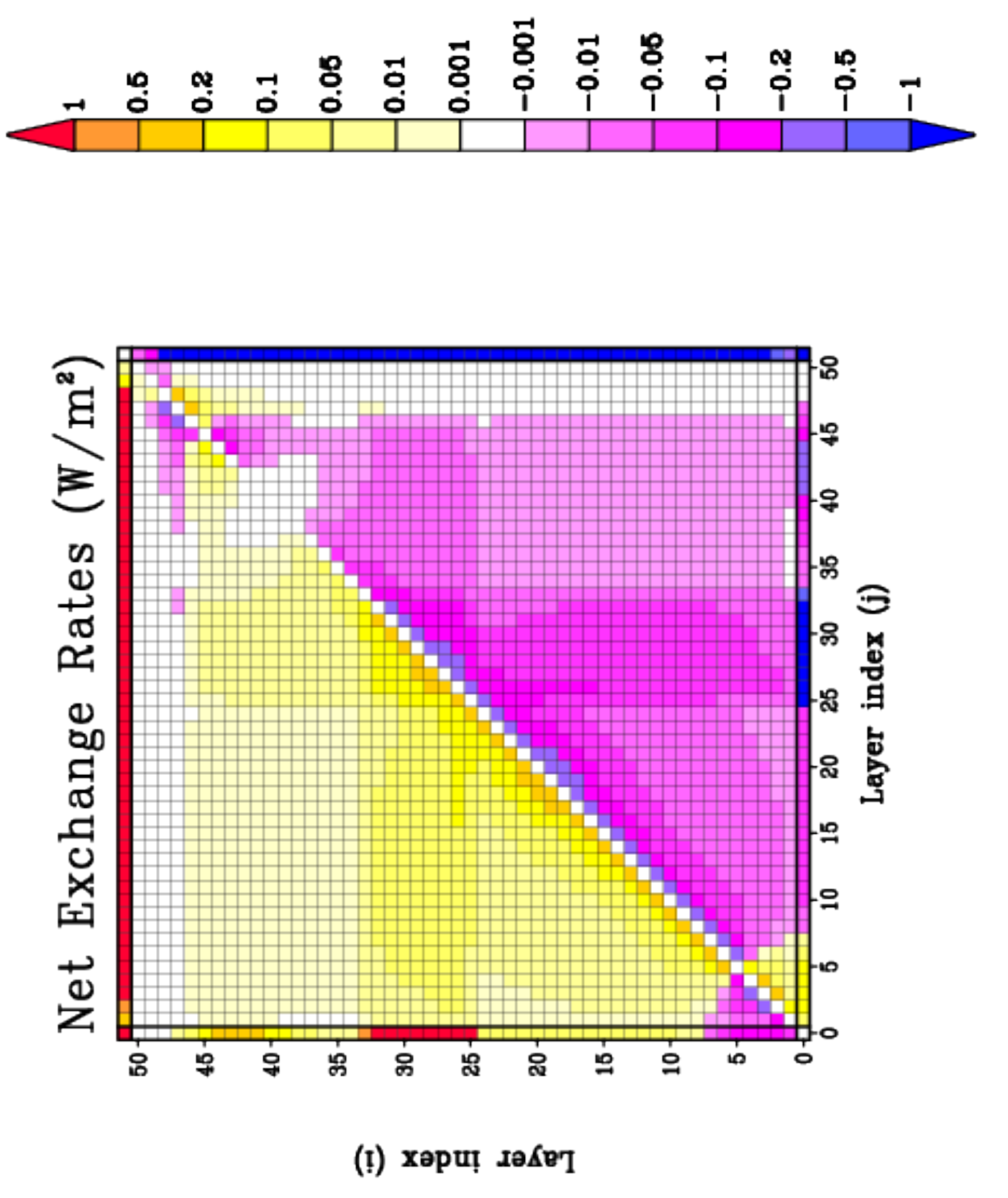,width=0.40\textwidth}\end{turn}}
}
    \caption{(a): heating rate ($K/day$) for MLS high-cloud configuration (ice cloud extending from $10.0$ to $12.0 km$ (layers $39-41$) with $IWC=0.0048 g.m^{-3}$ and $D_{e}=41.5 \mu m$)~; (b): same as (a) for SAW atmosphere (water cloud extending from $6.0$ to $8.0 km$ (layers $25-33$) and same $IWC$ and $D_{e}$)~; (c): NER matrix ($W/m^{2}$) in the high-cloud configuration for a MLS atmospheric profile~; (d): same as (c) for a SAW atmospheric profile.}
\label{fig:hc_ner}
\end{figure}

\begin{figure}[htbp]
\centering
\mbox{
    \subfigure[MLS all-clouds heating rate]{\begin{turn}{-90}\epsfig{figure=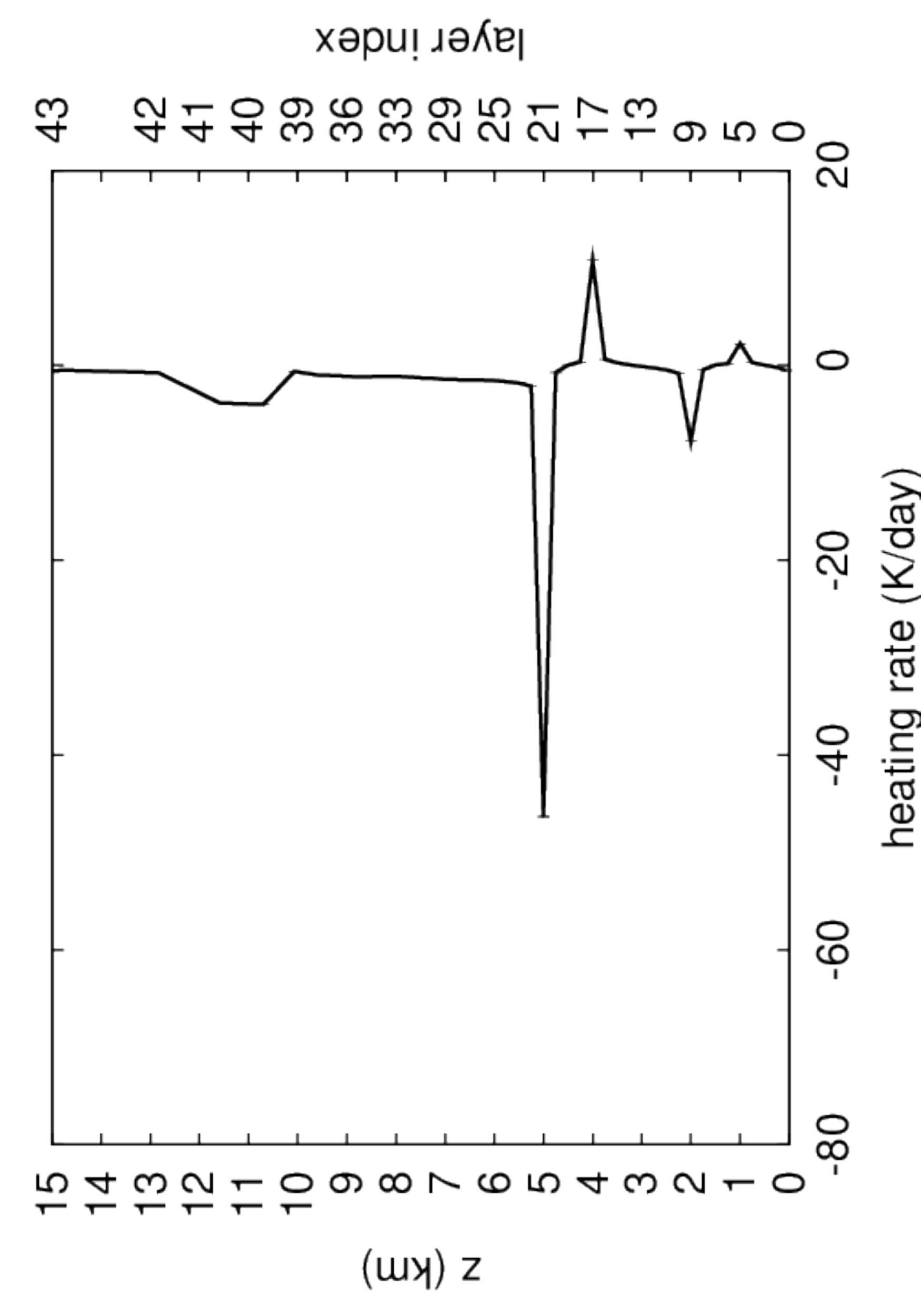,width=0.36\textwidth}\end{turn}}\quad
    \subfigure[SAW all-clouds heating rate]{\begin{turn}{-90}\epsfig{figure=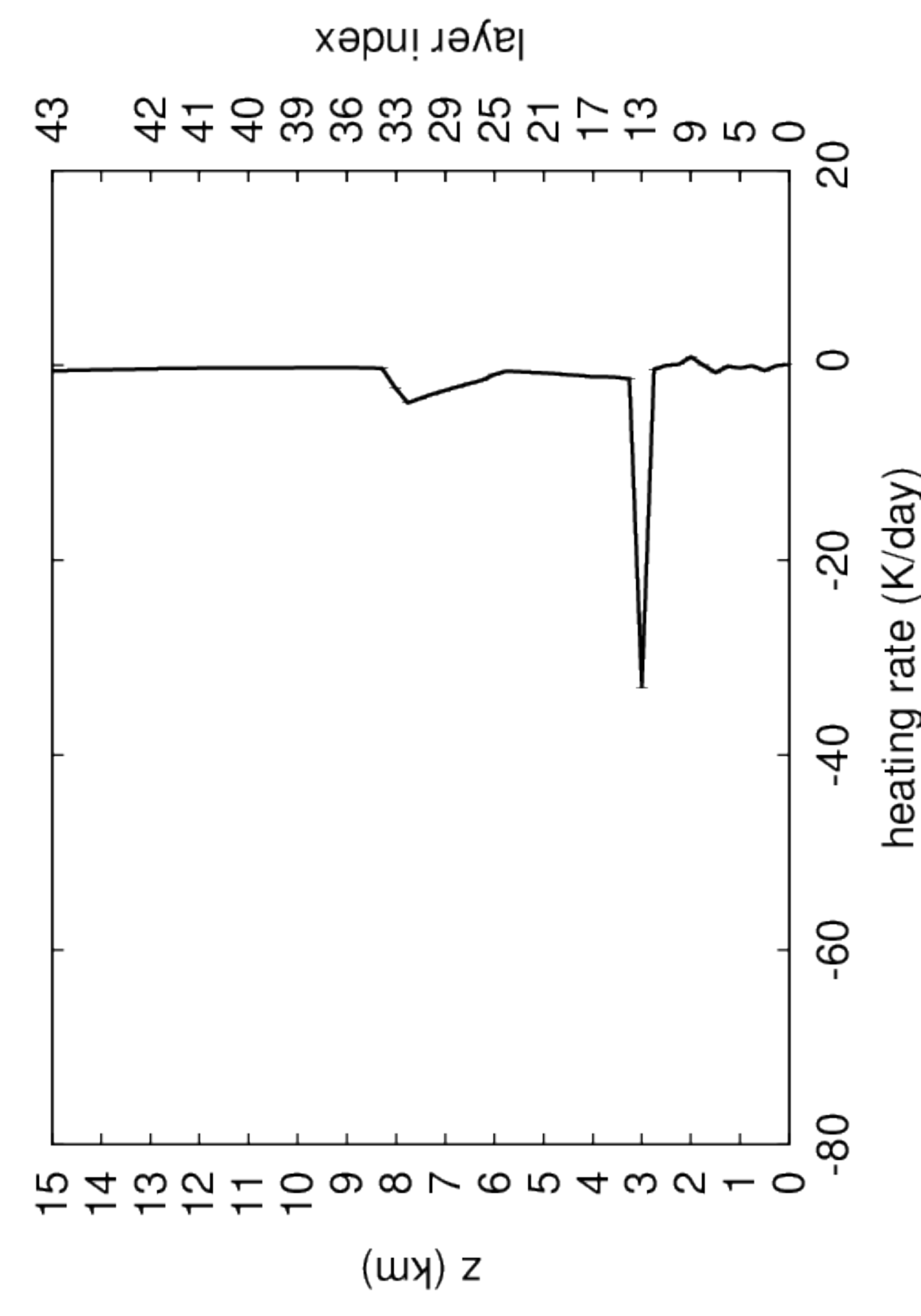,width=0.36\textwidth}\end{turn}}
}
\mbox{
    \subfigure[MLS all-clouds NER matrix]{\begin{turn}{-90}\epsfig{figure=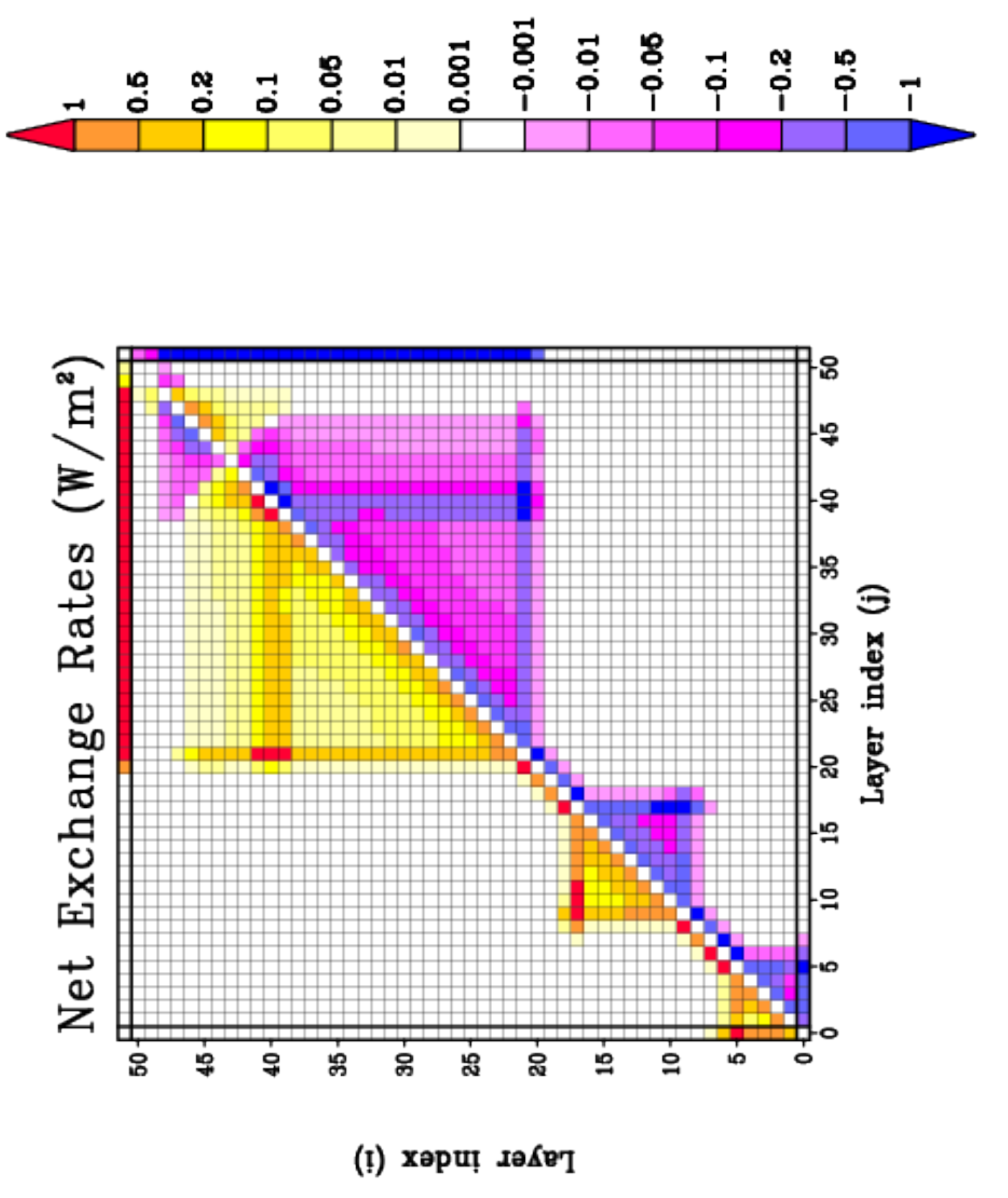,width=0.40\textwidth}\end{turn}}\quad
    \subfigure[SAW all-clouds NER matrix]{\begin{turn}{-90}\epsfig{figure=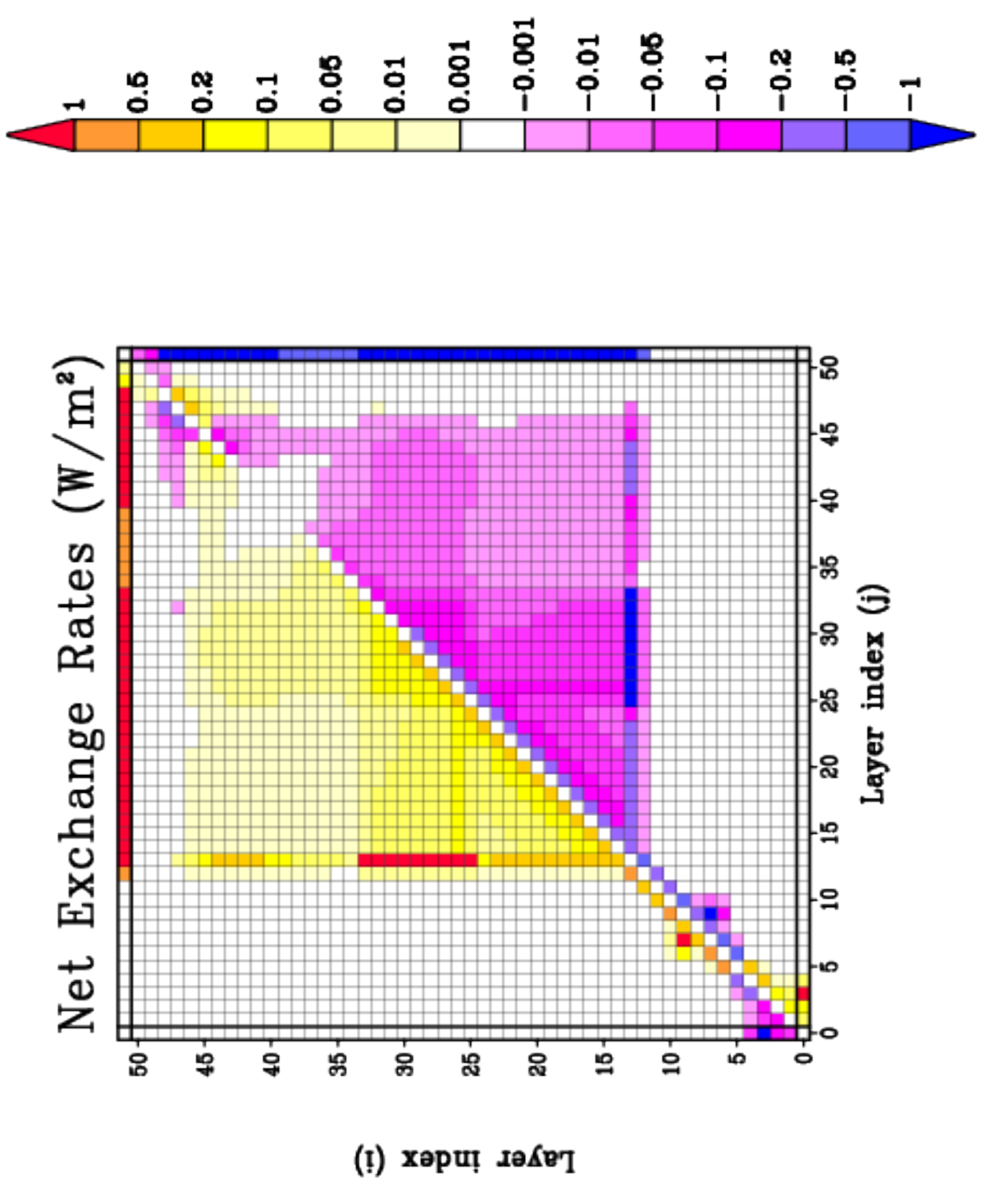,width=0.40\textwidth}\end{turn}}
}
    \caption{(a): heating rate ($K/day$) for MLS all-clouds configuration~; (b): same as (a) for SAW atmosphere~; (c): NER matrix ($W/m^{2}$) for a MLS all-clouds configuration~; (d): same as (c), for a SAW atmosphere.}
\label{fig:ac_ner}
\end{figure}

\fig{mc_ner}- \fig{ac_ner} present heating rates in $K/day$, in both (a) mid-latitude summer and (b) subarctic winter atmosphere, for cloudy configurations. Generally speaking, the bottom of all kind of clouds is heated, because the ground is warmer than the gas at cloud altitude. However, this is not true in the case of a low cloud with a SAW atmospheric profile: the  bottom of the cloud is located at an altitude $\approx 1 km$, which is the altitude of maximum temperature (higher than ground temperature). In all cases, the top of the cloud is cooled because of radiative exchanges with above gas layers and space, which have lower temperatures than the cloud. For optically thick clouds, such as a low and middle clouds in a MLS configuration, it can be seen that the middle of the cloud has a very small heating rate. In such cases, the middle layers are radiatively isolated from the rest of the system because of the high optical thickness of the cloud.

\fig{mc_ner}- \fig{ac_ner}(c) and (d) present the NER matrices in $W/m^{2}$ for the same configurations. For low clouds (figure not shown) and middle clouds (\ref{fig:mc_ner}) (with visible optical depths respectively $60$ and $72$), it can be seen that dominant exchanges for layers located below the cloud are between each atmospheric layer and the ground, between each atmospheric layer and the bottom of the cloud, and between close atmospheric layers (near diagonal NERs); similarly, for layers located above the cloud, dominant exchanges are between each layer and space, between each layer and the top of the cloud, and also between close atmospheric layers. In these two cases, the optical thickness of the cloud is high enough so that no radiative exchange can occur between atmospheric layers located on opposites sides of the cloud. For optically thick clouds, the cloudy layer behaves as a boundary (such as ground and space). The heating rate profile gives the same quantitative result: the bottom of the cloud is heated by the ground, and the top of the cloud is cooled by space in low and middle cloud configurations.

In the case of the high cloud (\fig{hc_ner}), which is an ice cloud with a visible optical depth of approximatively $0.8$, radiative exchange can occur between layers located below and above the cloud. The optical thickness of the cloud is not high enough for total extinction to occur. Dominant exchanges are between each layer and the ground, between each layer and space, between close atmospheric layers, and also between each cloud-free layer and cloudy atmospheric layers.

Finally, in the configuration where all clouds are present (\fig{ac_ner}), optically thick clouds (low and middle) behave as opaque screens, thus defining independent sections of the atmosphere which won't interact through radiation. Three main sections appear: the first, located below the low cloud~; the second, between low and middle clouds~; and the last, above the middle cloud.  On the other hand, optically thin clouds (high clouds) allow radiative transfer between layers located below and above these clouds.

In each section of the atmosphere, dominant exchanges are between cloud-free layers and the bottom boundary (ground or top of cloud), between cloud-free layers and the top boundary (space or bottom of cloud), and between adjacent cloud-free layers or, in the specific case of high clouds, between cloud-free layers and all cloudy layers.

\subsection{Spectral analysis}

For clear sky configurations (see \para{sec2}), it was chosen to present the spectral radiative budget $\Psi_{i,\nu}^{total}$ and its three components: $\Psi_{i,\nu}^{total}=\Psi_{i,\nu}^{gas-ground}+\Psi_{i,\nu}^{gas-space} +\Psi_{i,\nu}^{gas-gas}$. In this section, the same decomposition of spectral radiative budgets has been kept.

\begin{figure}[htbp]
\centering
\mbox{
    \subfigure[MLS middle cloud $\Psi_{i,\nu}^{total}$]{\begin{turn}{-90}\epsfig{figure=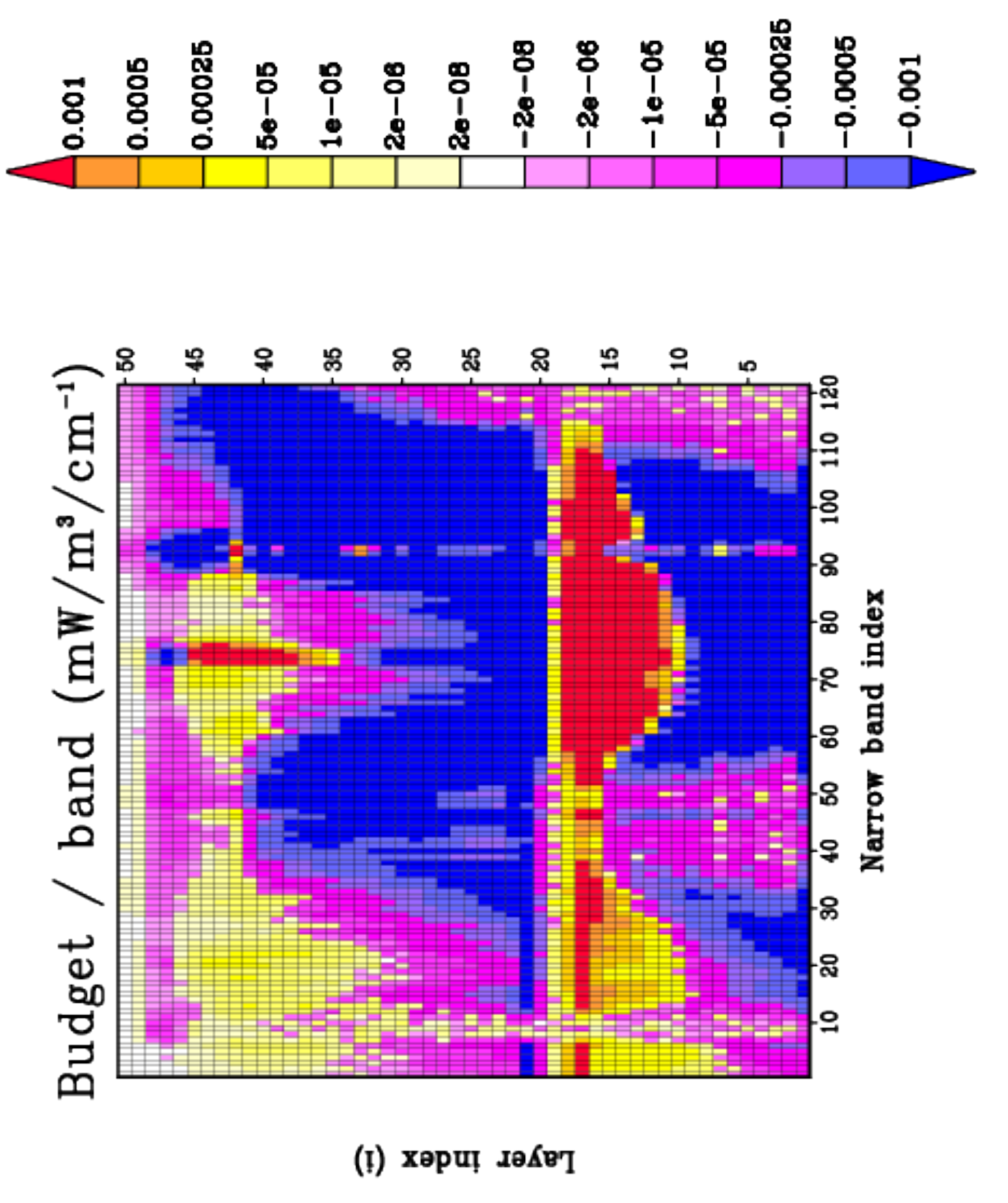,width=0.40\textwidth}\end{turn}}\quad
    \subfigure[MLS middle cloud $\Psi_{i,\nu}^{gas-ground}$]{\begin{turn}{-90}\epsfig{figure=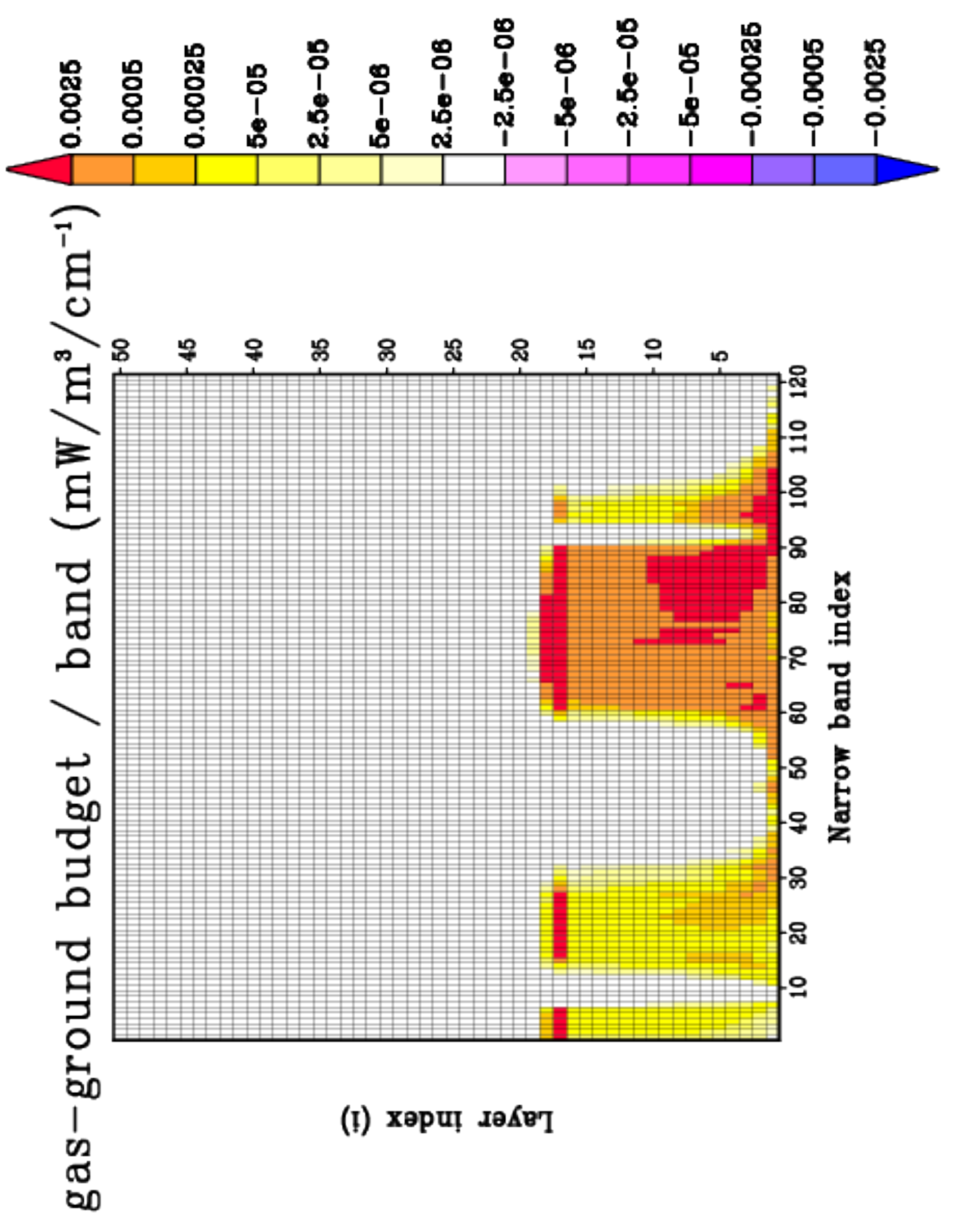,width=0.40\textwidth}\end{turn}}
}
\mbox{
    \subfigure[MLS middle cloud $\Psi_{i,\nu}^{gas-space}$]{\begin{turn}{-90}\epsfig{figure=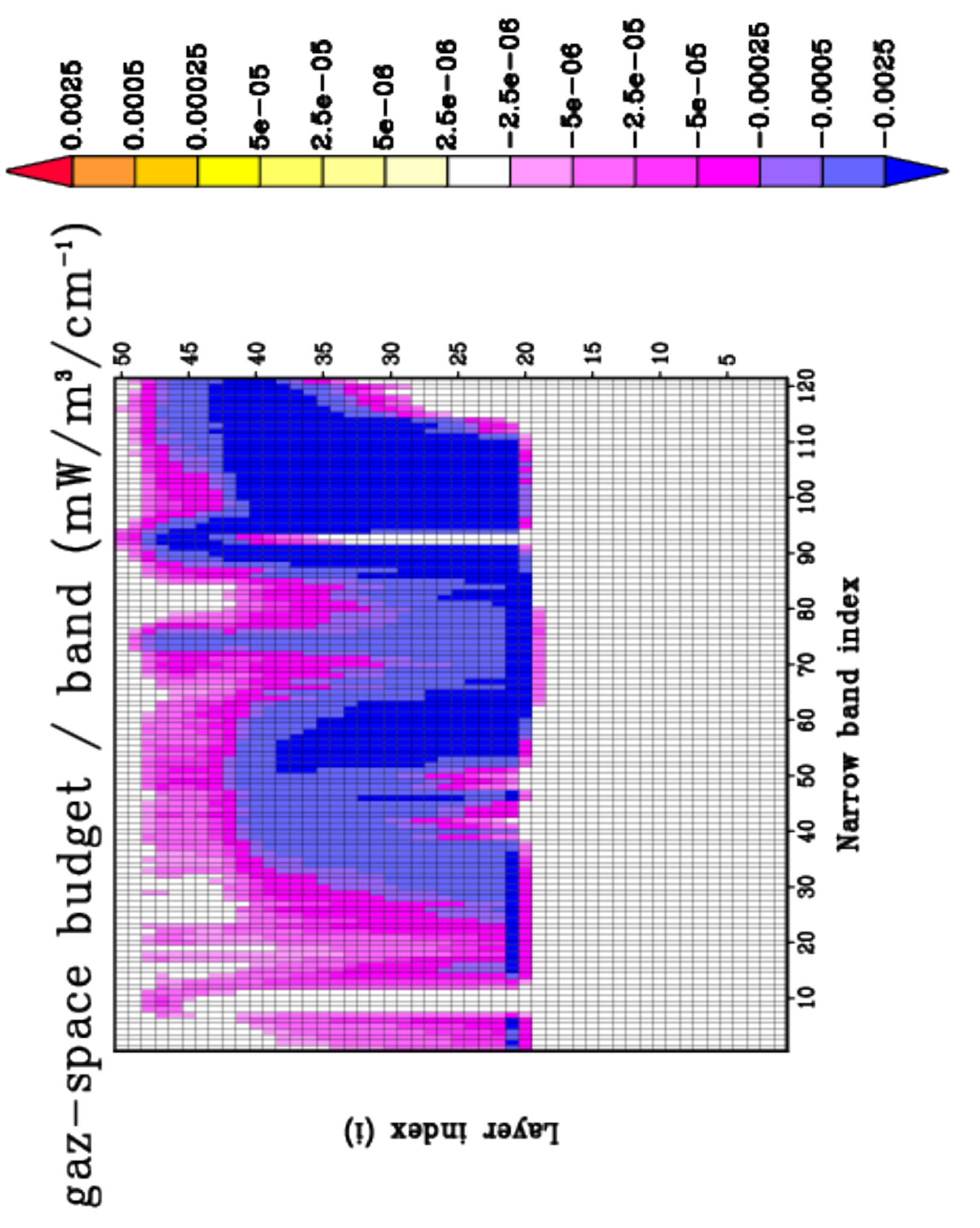,width=0.40\textwidth}\end{turn}}\quad
    \subfigure[MLS middle cloud $\Psi_{i,\nu}^{gas-gas}$]{\begin{turn}{-90}\epsfig{figure=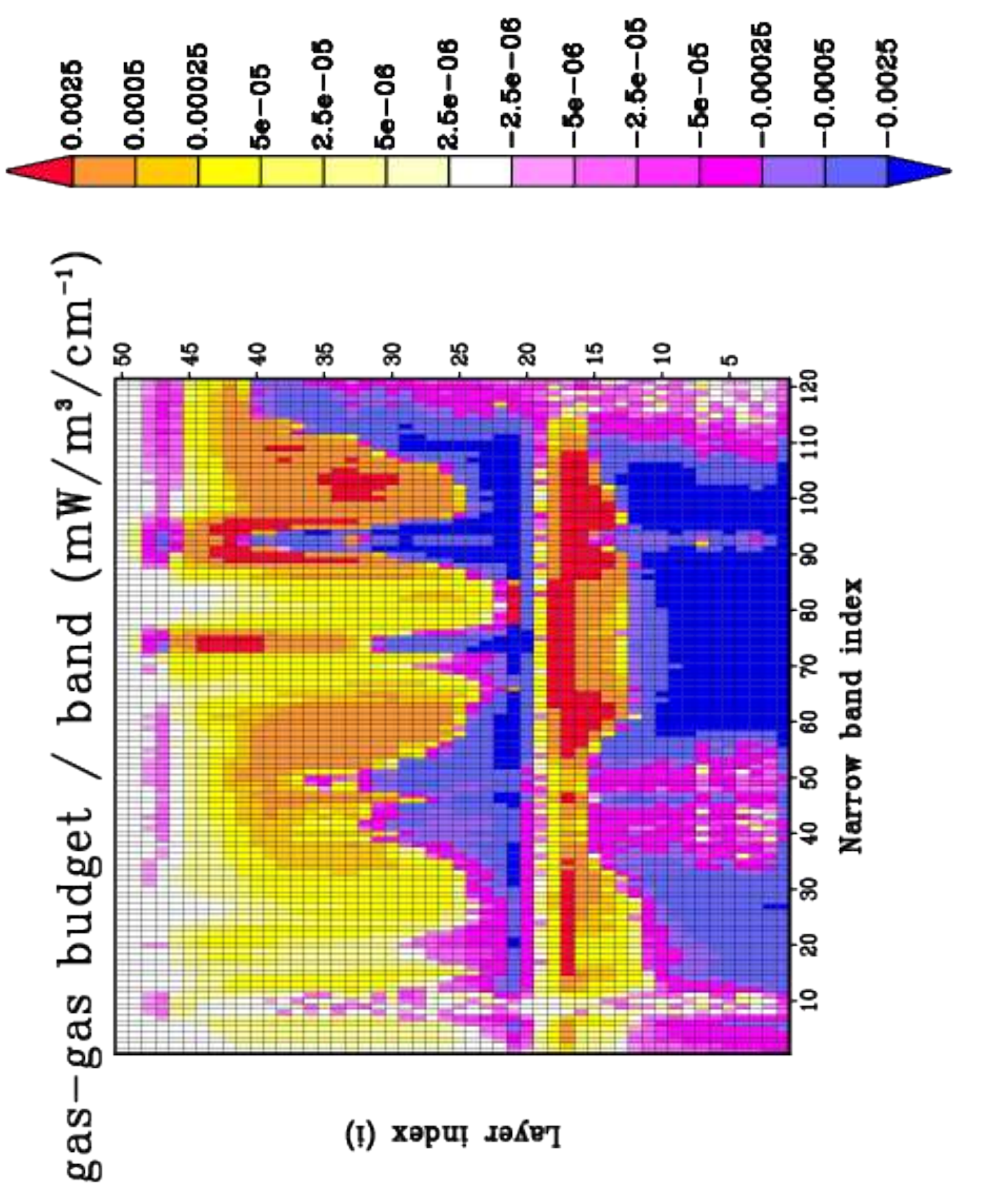,width=0.40\textwidth}\end{turn}}
}
    \caption{(a): total radiative budget ($mW/m^{3}/cm^{-1}$) as a function of narrow-band index (narrow-band width=20 $cm^{-1}$, from $4$ to $100 \mu m$) and atmospheric layer index, for middle cloud MLS configuration~; (b): net exchange between each atmospheric layer and ground ($mW/m^{3}/cm^{-1}$)~; (c): net exchange between each atmospheric layer and space ($mW/m^{3}/cm^{-1}$)~; (d): net exchange between each atmospheric layer and the rest of the atmosphere}
\label{fig:bil_mls_mc}
\end{figure}

NER matrices analysis in the previous section showed that optically thick clouds behave as opaque screens between layers located in either of their sides, thus dividing the atmosphere into radiatively independent sections. At the contrary, optically thin clouds behave as non opaque screen, allowing radiative transfers between layers located below and above such clouds. These were spectrally integrated results, and this section will now present the corresponding spectral decomposition of NERs in terms of net radiative budgets: see \fig{bil_mls_mc}.

These figures first show that the net exchanges between each atmospheric layer and ground, in the presence of optically thick clouds (\fig{bil_mls_mc}(b)), can occur only for layers located below the cloud, only in low absorption spectral regions. For high absorption spectral regions (such as the water or the $CO_{2}$ absorption bands), radiation emitted from the ground is totally absorbed within the first atmospheric layer, net exchanges between ground and higher layers being therefore impossible. Similarly, \fig{bil_mls_mc}(c) shows that net exchanges between each atmospheric layer and space can occur only above optically thick clouds, in low absorption spectral regions. This confirms that optically thick clouds behave as opaque screens between the ground and space. This is true over the whole infrared spectra, but this effect is only visible in low absorption spectral regions because net exchanges between boundaries and gaseous layers are only possible in these spectral regions.

Optically thin high clouds (figures not shown) do not behave as opaque screens: net exchanges are possible between ground and layers located above the cloud, and between space and layers located below the cloud, still for low absorption spectral regions.

\section{Effects of scattering}
\label{para:sec4}

The effects of the scattering process are generally neglected by the G.C.M. community in atmospheric longwave radiative transfer computations. However, increasing precision requirements \footnote{For instance ground and space radiative budgets need to be known with an uncertainty of less than one percent.} now imply that scattering should be accounted for in G.C.M. longwave radiative transfer schemes \cite{Dufresne02}. Within this framework, this section will present the effect of scattering for the same cloudy configurations as used in preceding sections. First of all, the effect of scattering will be shown over ground and space total radiative budgets. Then it will be shown over previously introduced NER matrices and heating rate profiles. Finally, these results will be spectrally decomposed in order to understand which longwave spectral ranges are affected by scattering. The results of this analysis are expected to be used in order to set up a new parameterization of longwave radiative transfers, that accounts for scattering, in an already existing Global Circulation Model.

\subsection{Effects of scattering on ground and space total radiative budgets}

Tables \ref{Ground} and \ref{Space} present the effect of scattering on the total radiative budgets of ground and space, for each cloudy configuration. Modification of ground total radiative budget due to scattering ranges from $1$ to $6 \%$, with a peak of $10 \%$ for the SAW middle cloud configuration. Statistical errors are never greater than $0.2\%$. As far as space is concerned (radiative flux toward space), scattering has a contribution of $1$ to $5 \%$ over the total radiative budget, with error estimates below $0.1 \%$.\footnote{These results are compatible with already published results \cite{Fu01}~; the purpose of this paper is not to present a strict comparison of results with other authors, and thus validate the analysis tool that we have developped. Its main purpose is rather to present a clear analysis of longwave radiative transfers, in terms of Net Exchange Rates, concerning the role that scattering may play in terrestrial atmospheric longwave radiative transfers.}

The fact that scattering should be considered as negligible or not, for longwave atmospheric radiative transfer computations, depends entirely on the required accuracy level: for rough estimations, the absorption approximation is suitable~; when a higher precision is required ($1\%$), scattering should be taken into consideration.

\begin{table}[ht] \begin{center}
\begin{tabular}{|c||c|c|c|}
\hline
\textbf{Configuration} & \textbf{$\Psi_{ground}$ A.A. ($W/m^{2}$) } & \textbf{$\Delta \Psi_{ground}$ ($W/m^{2}$)} & \textbf{$\Delta \Psi_{ground}$ $(\%)$ } \\ \hline
MLS low cloud & -8.89  & 0.317 $\pm$ 0.005 & -3.57 $\pm$ 0.05 \\ \hline
MLS middle cloud & -29.45  & 0.788 $\pm$ 0.01 & -3.68 $\pm$ 0.05 \\ \hline
MLS high cloud & -69.73  & 0.743 $\pm$ 0.04 & -1.07 $\pm$ 0.06 \\ \hline
MLS all clouds & -8.89  & 0.325 $\pm$ 0.005 & -3.65 $\pm$ 0.05 \\ \hline
SAW low cloud & 2.51  & -0.16 $\pm$ 0.0009 & -6.27 $\pm$ 0.03 \\ \hline
SAW middle cloud & -1.07  & 0.11 $\pm$ 0.003 & -10.67 $\pm$ 0.2 \\ \hline
SAW high cloud & -71.00  & 2.38 $\pm$ 0.05 & -3.35 $\pm$ 0.07 \\ \hline
SAW all clouds & -2.51  & -0.12 $\pm$ 0.0009 & -5.00 $\pm$ 0.03 \\ \hline
\end{tabular}
\end{center}
\myKaption{Effect of scattering on ground radiative budget. First column presents the cloud configuration. Second column presents the ground total radiative budget $\Psi_{ground}$ in $W/m^{2}$ computed using the Absorption Approximation (A.A), \cite{Li01}. The third column presents the effect of scattering on ground total radiative budget $\Delta \Psi_{ground}$ in $W/m^{2}$ computed as the difference between the radiative budget taking into account scattering and the radiative budget with the absorption approximation. The reported uncertainties have been estimated from the Monte Carlo statistical standard deviations. The fourth column shows $\Delta \Psi_{ground}$ expressed in terms of percentage, with its statistical uncertainty.}
\label{Ground}
\end{table}

\begin{table}[ht] \centering
\begin{tabular}{|c||c|c|c|} \hline
\textbf{Configuration} & \textbf{$\Psi_{space}$ A.A. ($W/m^{2}$) } & \textbf{$\Delta \Psi_{space}$ ($W/m^{2}$)} & \textbf{$\Delta \Psi_{space}$ $(\%)$ } \\ \hline
MLS low cloud & 286.71  & -4.12 $\pm$ 0.16 & -1.44 $\pm$ 0.05 \\ \hline
MLS middle cloud & 247.18  & -5.50 $\pm$ 0.10 & -2.22 $\pm$ 0.04 \\ \hline
MLS high cloud & 238.20  & -10.38 $\pm$ 0.17 & -4.36 $\pm$ 0.07 \\ \hline
MLS all clouds & 207.81 & -10.36 $\pm$ 0.11 & -4.99 $\pm$ 0.05 \\ \hline
SAW low cloud & 210.18 & -4.09 $\pm$ 0.10 & -1.95 $\pm$ 0.05 \\ \hline
SAW middle cloud & 200.25  & -4.63 $\pm$ 0.08 & -2.31 $\pm$ 0.04 \\ \hline
SAW high cloud & 182.04  & -7.74 $\pm$ 0.12 & -4.25 $\pm$ 0.06 \\ \hline
SAW all clouds & 176.52  & -8.37 $\pm$ 0.095 & -4.74 $\pm$ 0.05 \\ \hline
\end{tabular}
\myCaption{Same as table \ref{Ground} for space radiative budget}
\label{Space}
\end{table}

\subsection{Effects of scattering on NER and heating rates}

\begin{figure}[htbp]
\centering
\mbox{
\subfigure[]{\begin{turn}{-90}\epsfig{figure=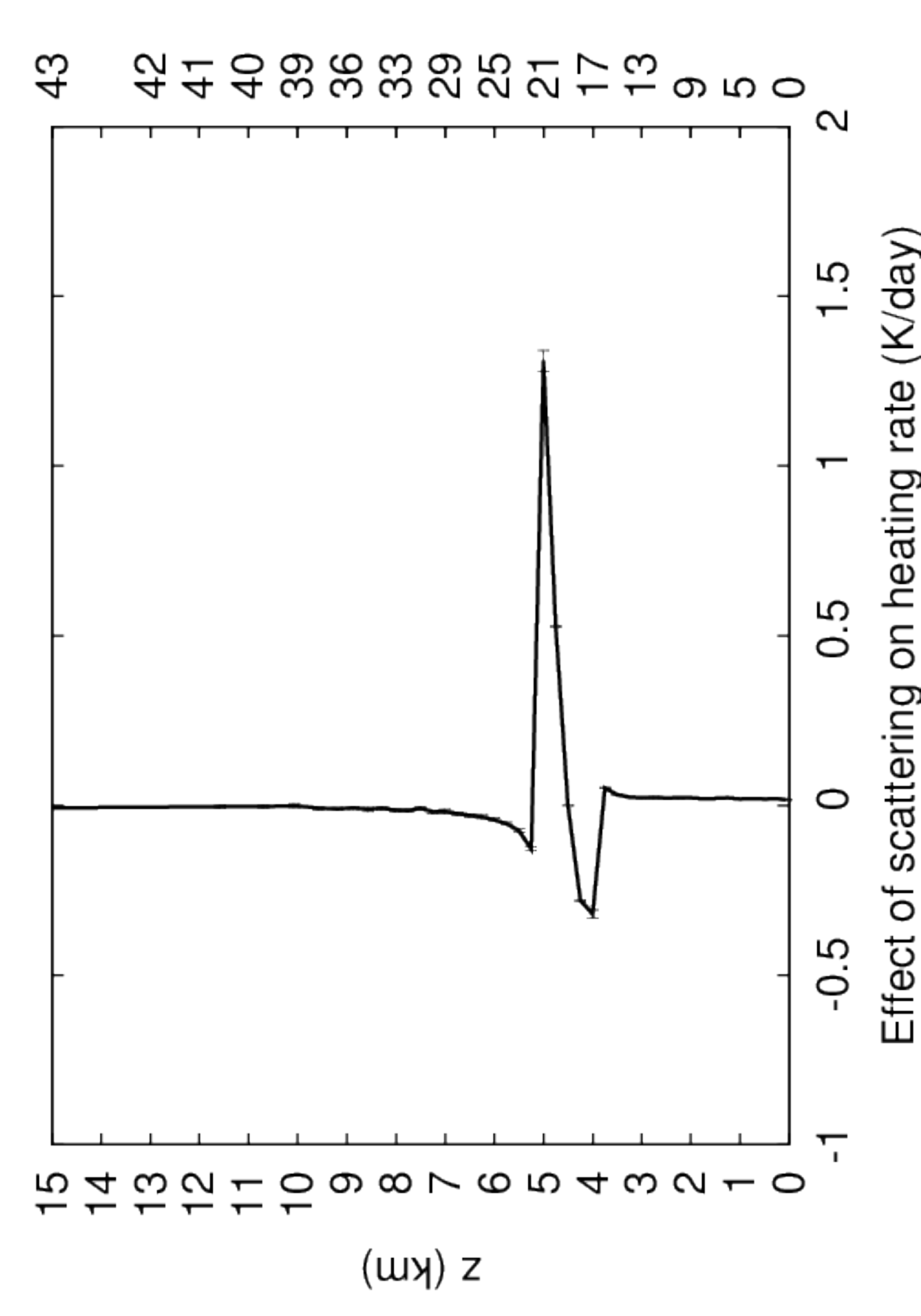,width=0.36\textwidth}\end{turn}}\quad
\subfigure[]{\begin{turn}{-90}\epsfig{figure=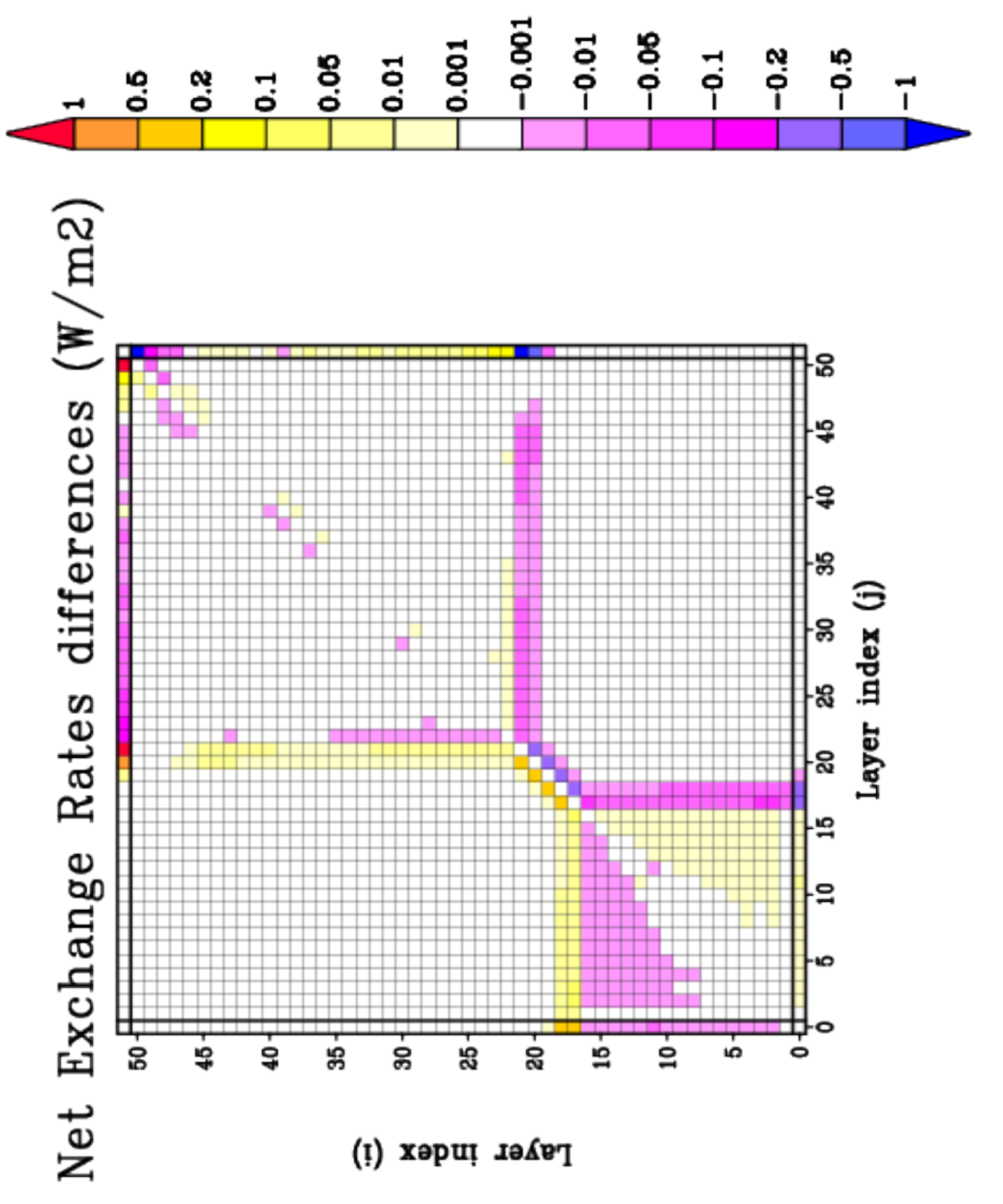,width=0.40\textwidth}\end{turn}}
}
    \caption{(a) effect of scattering ($mW/m^{3}$) on MLS middle-cloud heating rates~; (b) effect of scattering ($W/m^{2}$) on MLS middle-cloud NER matrix.}
\label{fig:diff_mls_mc}
\end{figure}

\begin{figure}[htbp]
\centering
\mbox{ 
\subfigure[]{\begin{turn}{-90}\epsfig{figure=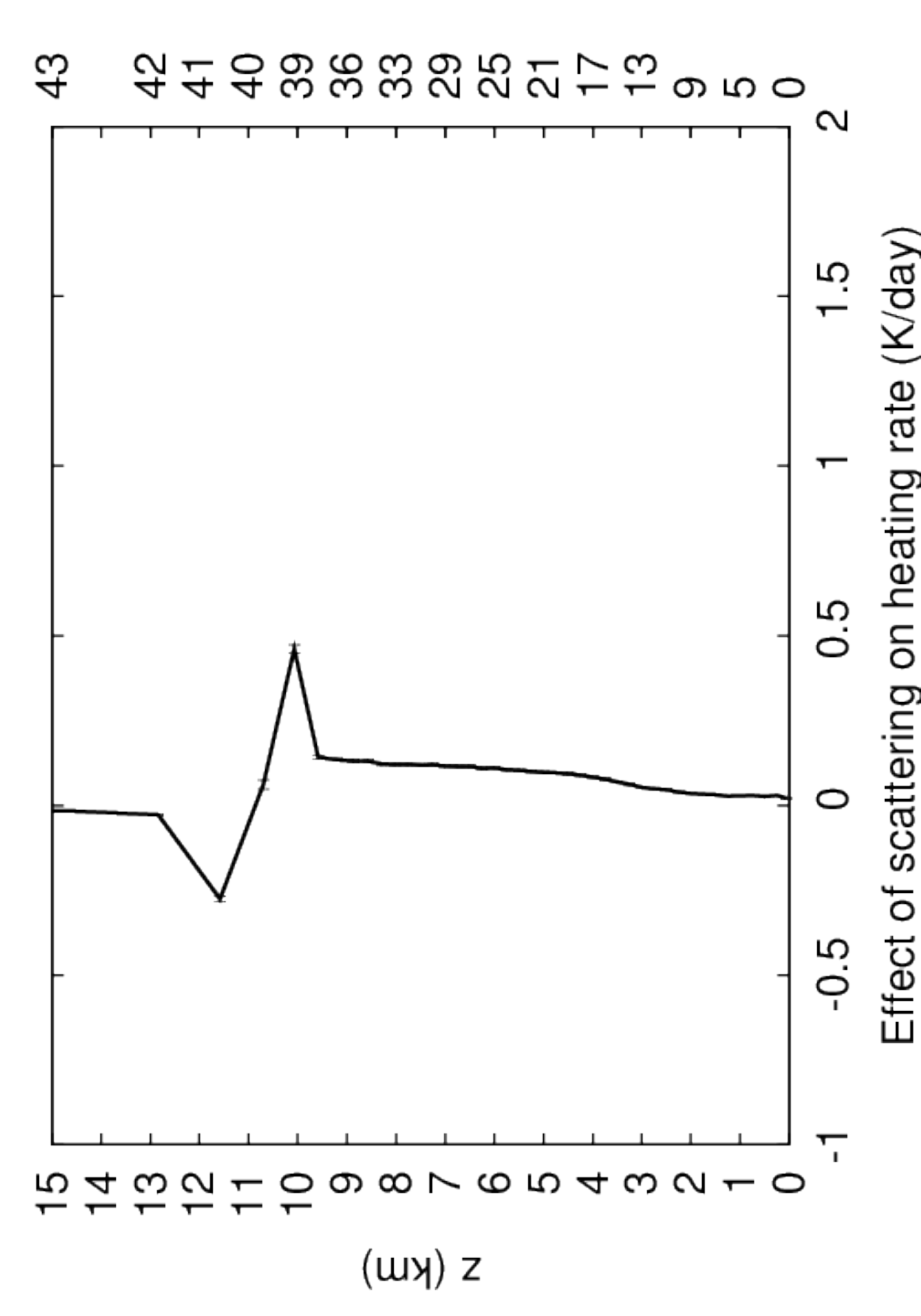,width=0.36\textwidth}\end{turn}}\quad
\subfigure[]{\begin{turn}{-90}\epsfig{figure=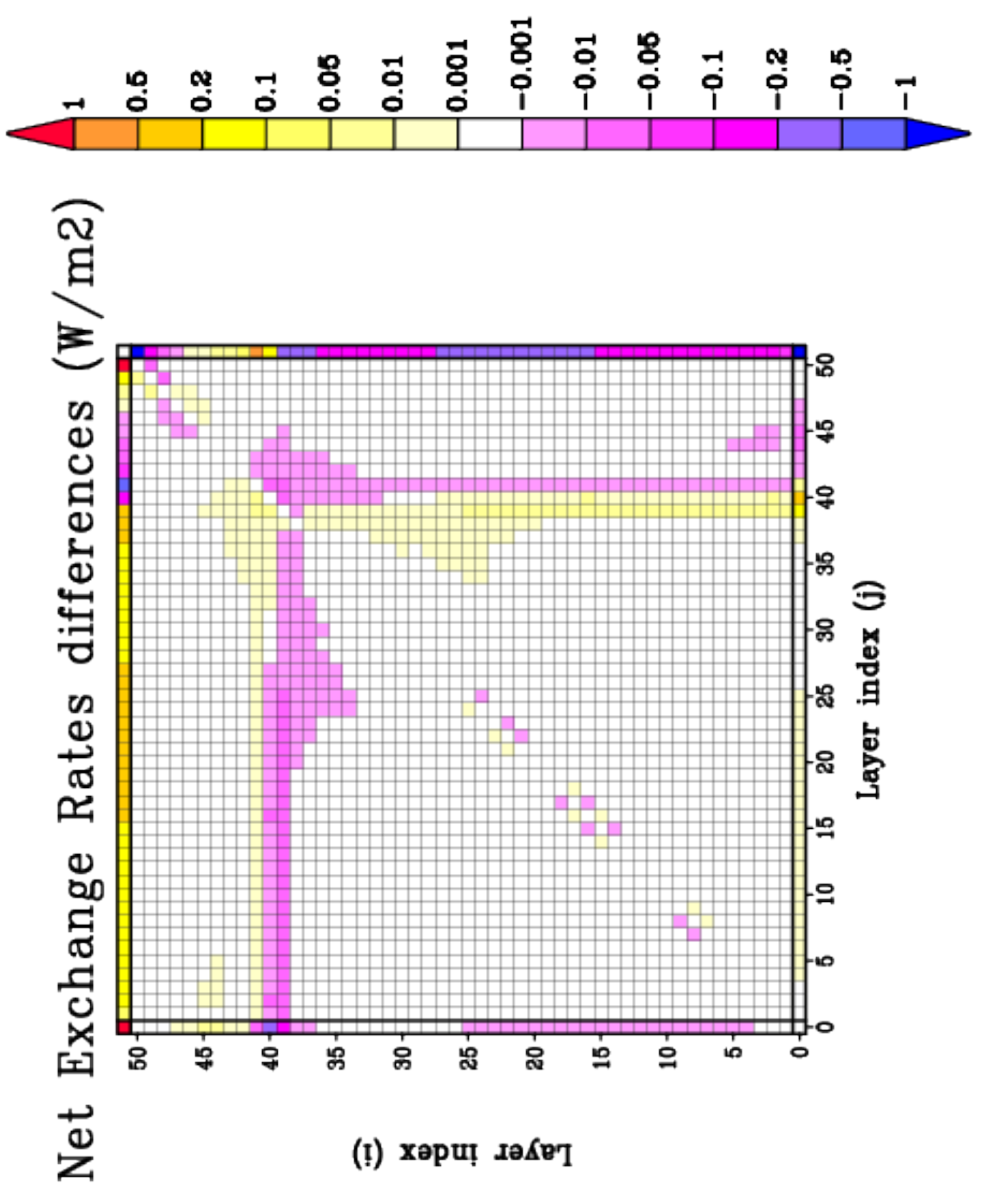,width=0.40\textwidth}\end{turn}}
}
    \caption{(a) effect of scattering ($mW/m^{3}$) on MLS high-cloud heating rates~; (b) effect of scattering ($W/m^{2}$) on MLS high-cloud NER matrix.}
\label{fig:diff_mls_hc}
\end{figure}

\fig{diff_mls_mc} and \fig{diff_mls_hc} display the effect of scattering on the heating rates for MLS middle and high cloud configurations. These results have been computed as the differences of NERs between a first computation for which scattering has been taken into account, and a second computation, for which scattering has been neglected (using the Absorption Approximation, see \cite{Fu01}). These results show that scattering mainly affects the clouds edges, and especially the top of the clouds. This effect is small for low and middle clouds: it is no higher that 2$\%$ of the absolute value of the heating rate at clouds edges (\fig{diff_mls_mc}(a) compared to \fig{mc_ner}(a)). For those clouds, the effect of scattering is to decrease the absolute value of the heating rate at edges, i.e. to decrease the apparent absorption of the cloud.

For high and optically thin clouds, the effect of scattering on heating rate is not negligible and is as high as 10$\%$ at the cloud edges (\fig{diff_mls_hc}(a) compared to \fig{hc_ner}(a)). In this configuration, scattering now increases the absolute value of the heating rate at cloud edges. However, scattering is still decreasing the absolute value of the total absorption by the cloud.

Let us try to understand these results, from simple considerations based on the following assumptions: radiation incident at the surface of clouds $I$ will be assumed isotropic, no matter what are the sources. Clouds are also considered as isothermal, which is fairly well justified for radiative exchanges between a cloud and ground or space. Within these assumptions, the intensity of the exchange between radiative sources and the cloud is proportional to $I$, and its expression is given by equation \ref{eq:I} below:

\begin{equation}
I \int_{a}^{+\infty}[1-exp(-k_{a}l)]p(l)dl
\label{eq:I}
\end{equation}

In the above equation, $k_{a}$ is the absorption coefficient of the cloud and $l$ is the length of scattering paths within the cloud. $l$ is distributed according to the probability density function $p$ on $[a,+\infty[$ where $a$ is the smallest path length that can be encountered through the medium.

For a purely absorbing cloud ($k_{s}=0$), the smallest path length within the cloud is $a=e$, where $e$ is the cloud width. In this case, the cloud behaves as a blackbody at the optically thick limit for absorption ($k_{a}e>>1$).

If scattering is taken into account ($k_{s} \neq 0$), very small paths must be accounted for ($a=0$). In this case, the cloud no longer behaves as a blackbody even for large $k_{a}$ values. The mean path length $<l>$ remains a constant no matter what are the scattering properties of the cloud \cite{Blanco01}, but the effect of scattering is to create a lot of short paths and some very long paths: scattering widens the path length distribution (increasing the standard deviation of $l$). In the case of optically thin clouds for absorption, this has no effect as $\int_{0}^{+\infty}[1-exp(-k_{a}l)]p(l)dl \approx k_{a}<l>$ is constant when the cloud scattering properties change: for such clouds, scattering will not introduce any significant modifications of the cloud total radiative budget. However, for clouds which are not optically thin for absorption (which is the the case most of the time for Earth configurations), convexity properties of the exponential function will result  in a net decrease of $\int_{0}^{+\infty}[1-exp(-k_{a}l)]p(l)dl$, thus of the cloud total radiative budget. This is what was found for Earth cloudy configurations:

\begin{itemize}

\item For low (figure not shown) and middle cloud (\fig{diff_mls_mc}) configurations, the cloud total radiative budget is negative, and scattering reduces its absolute value. The two maxima in the heating rate profile (at the bottom and at the top of the cloud) are both attenuated by scattering.

\item For high cloud configurations (\fig{diff_mls_hc}), the cloud total radiative budget is also negative. Even if scattering is to increase locally the heating rate peaks (radiative exchanges are concentrated at cloud edges), the heating rate at the top the cloud is more strongly attenuated than it is increased at the bottom of the cloud. This results in a net decrease of the absolute value of the cloud total radiative budget.

\end{itemize}

\subsection{Spectral analysis}

\begin{figure}[htbp]
\centering
\mbox{
    \subfigure[$\delta \Psi_{i,\nu}^{total}$]{\begin{turn}{-90}\epsfig{figure=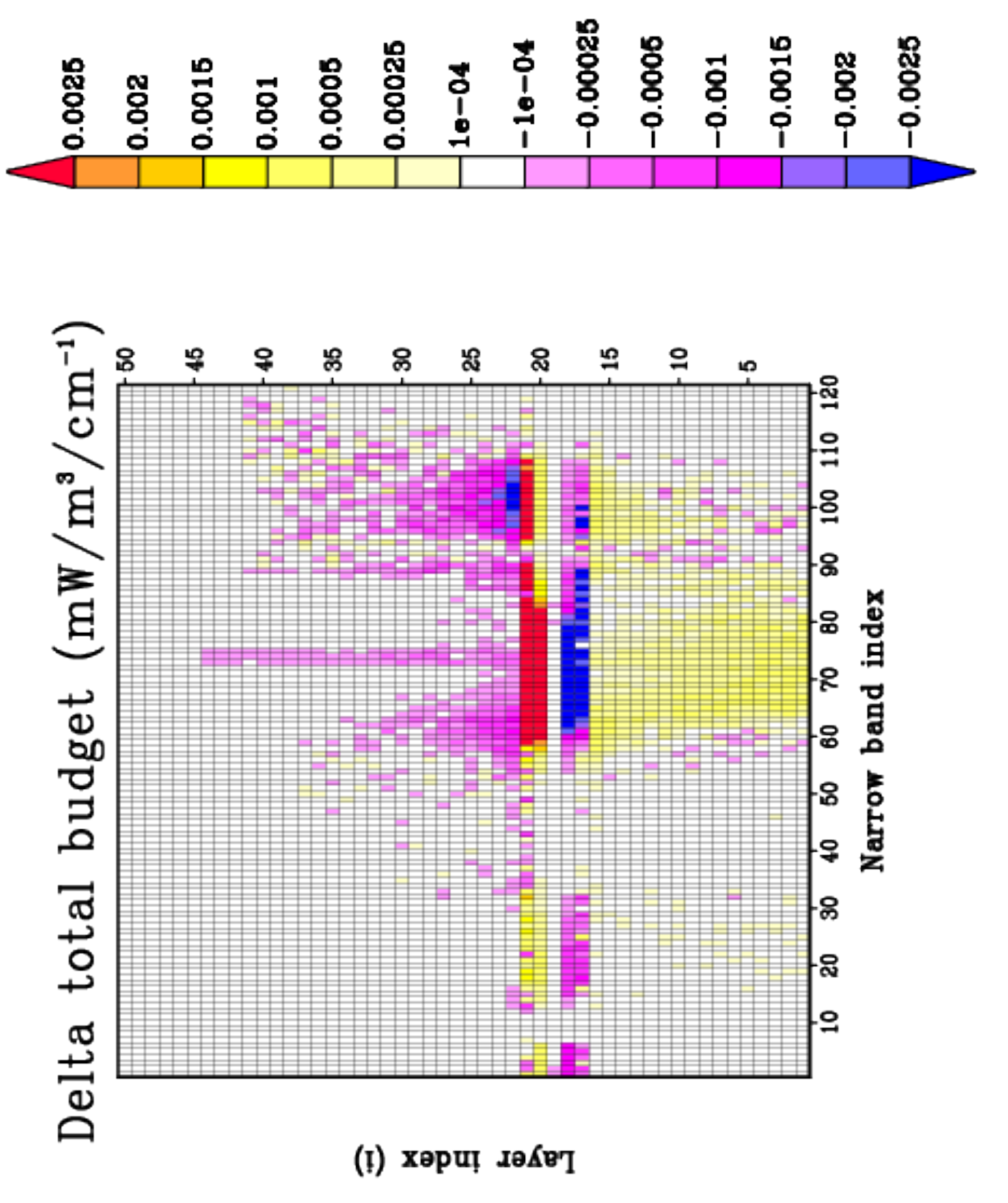,width=0.40\textwidth}\end{turn}}\quad
    \subfigure[$\delta \Psi_{i,\nu}^{gas-ground}$]{\begin{turn}{-90}\epsfig{figure=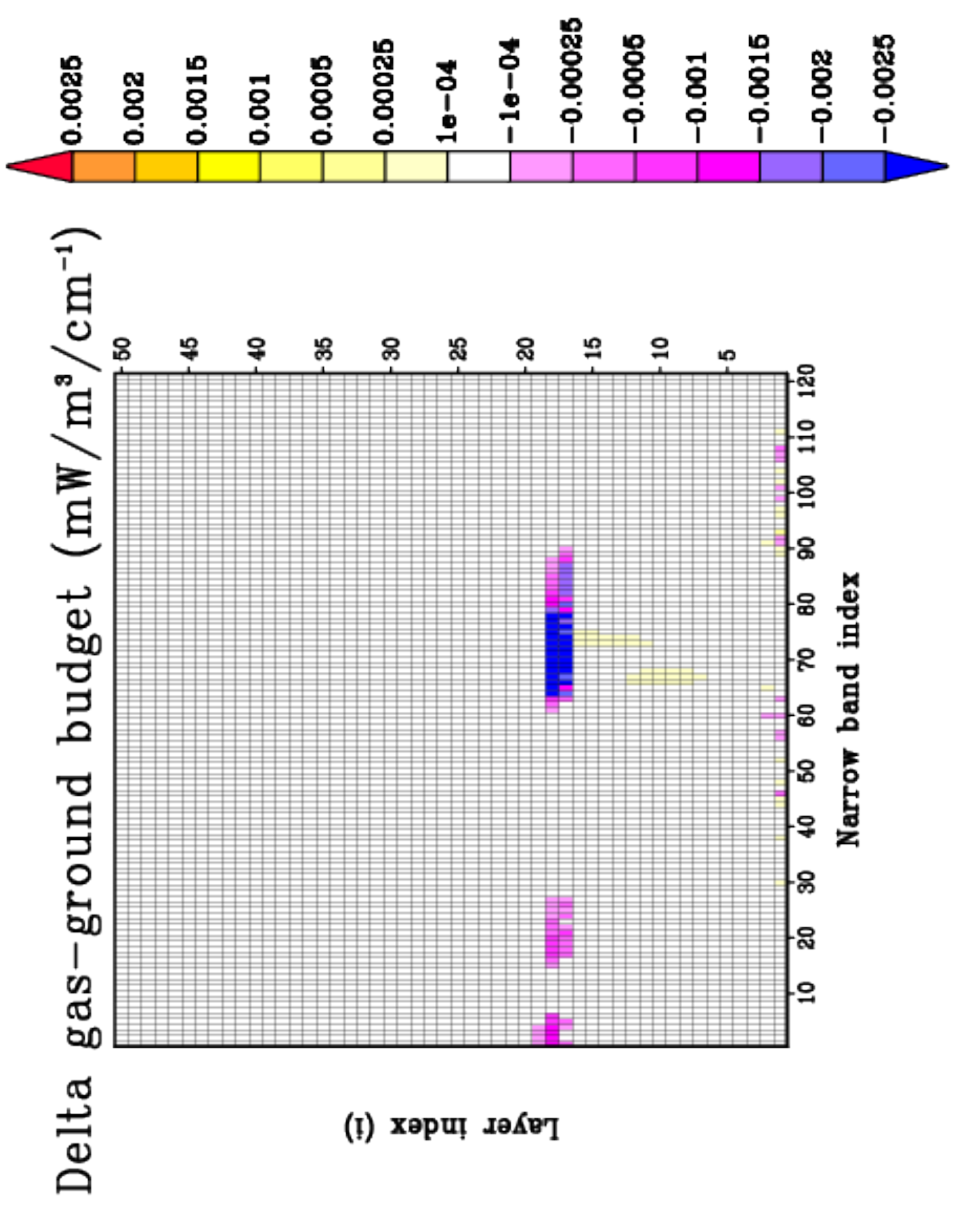,width=0.40\textwidth}\end{turn}}
}
\mbox{
    \subfigure[$\delta \Psi_{i,\nu}^{gas-space}$]{\begin{turn}{-90}\epsfig{figure=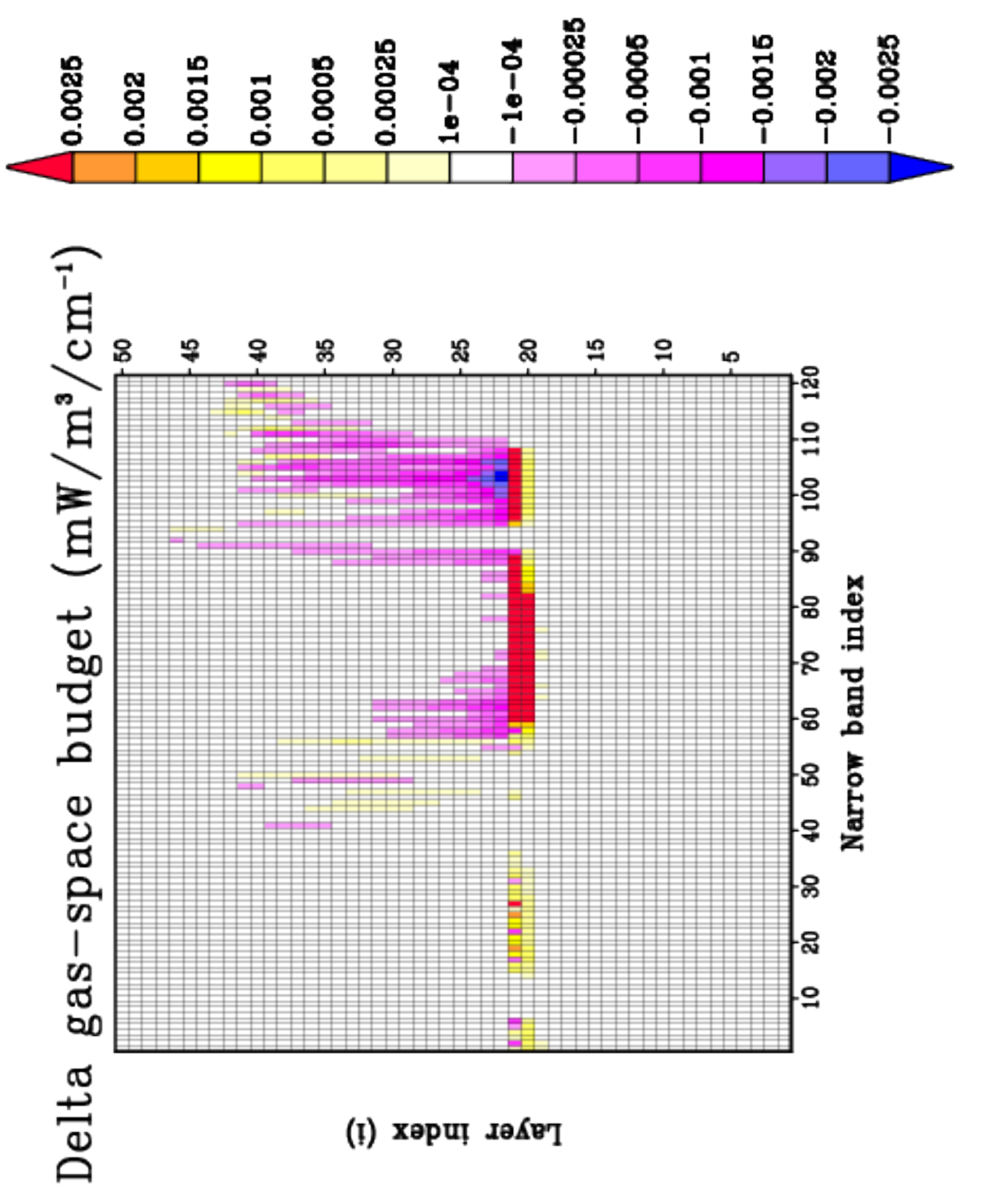,width=0.40\textwidth}\end{turn}}\quad
    \subfigure[$\delta \Psi_{i,\nu}^{gas-gas}$]{\begin{turn}{-90}\epsfig{figure=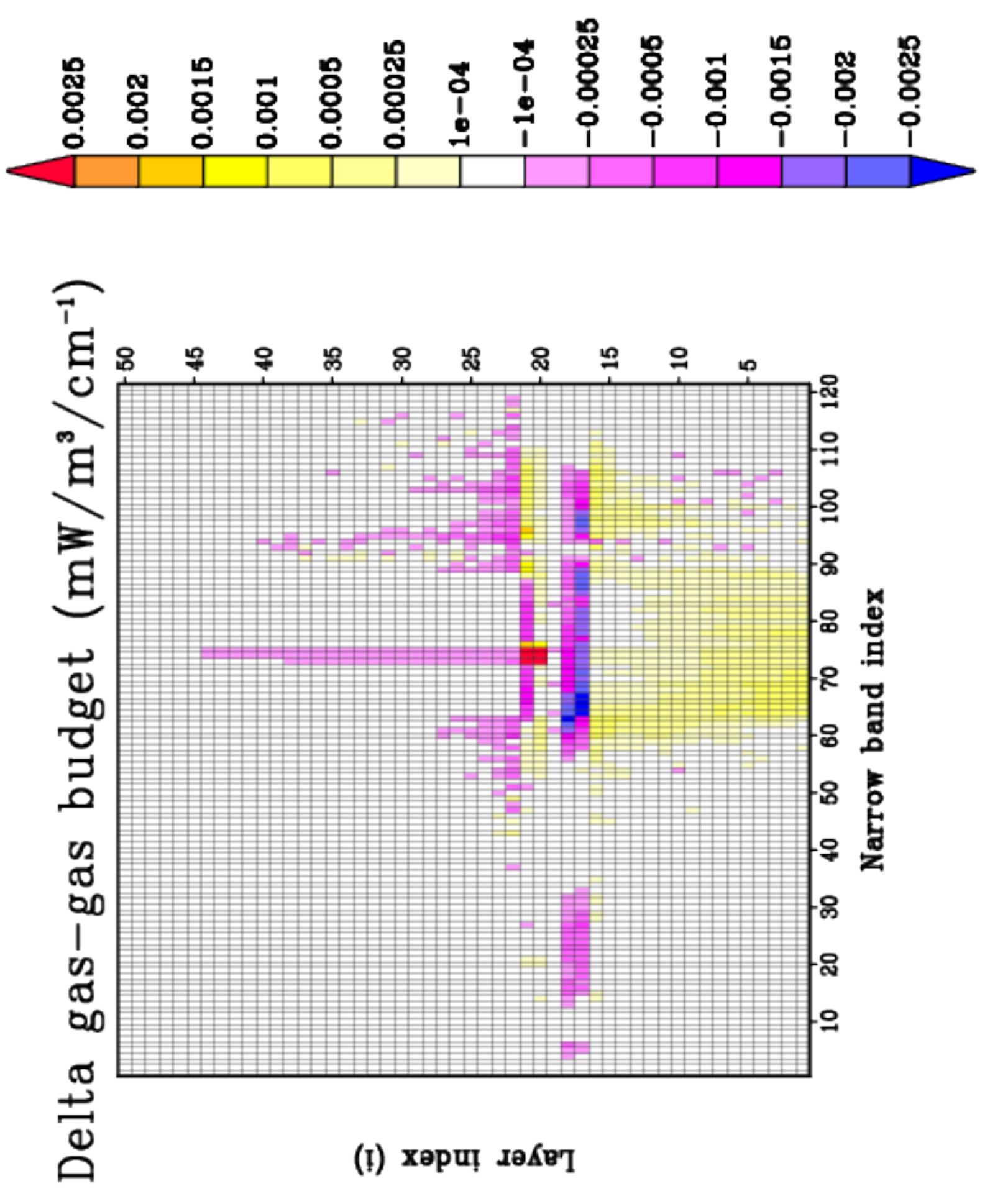,width=0.40\textwidth}\end{turn}}
}
    \caption{(a): effect of scattering on total radiative budget ($mW/m^{3}/cm^{-1}$) as a function of narrow-band index (narrow-band width=20 $cm^{-1}$, from $4$ to $100 \mu m$) and atmospheric layer index, for middle-cloud MLS configuration~; (b): effect of scattering on net exchange between each atmospheric layer and ground ($mW/m^{3}/cm^{-1}$)~; (c): effect of scattering on net exchange between each atmospheric layer and space ($mW/m^{3}/cm^{-1}$)~; (d): effect of scattering on net exchange between each atmospheric layer and the rest of the atmosphere}
\label{fig:diff_bil_mls_mc}
\end{figure}

\fig{diff_bil_mls_mc} represent the effect of scattering
($mW/m^{3}/cm^{-1}$) on radiative budget $\Psi_{i,\nu}^{total}$ and its
decomposition (net exchanges between each atmospheric layer and ground
$\Psi_{i,\nu}^{gas-ground}$, between each atmospheric layer and space
$\Psi_{i,\nu}^{gas-space}$, and between each atmospheric layer and the
rest of the atmosphere $\Psi_{i,\nu}^{gas-gas}$), for a MLS middle-cloud
configuration.

Conclusions about the effects of scattering on spectral heating rates are very similar to the previous results: scattering mainly modifies net exchanges between boundaries and cloudy layers. In low-cloud (figures not shown) and middle-cloud (\fig{diff_bil_mls_mc}) configurations, scattering decreases net exchanges between ground and cloudy layers, which means that photons emitted from the ground are backscattered by optically thick clouds~; in high-cloud configurations (figures not shown), scattering mainly increases these net exchanges, which means that optically thin clouds do not significantly backscatter photons emitted from the ground~; rather, photons do cross the cloud surface and scattering increases their mean path length, thus increasing the amount of energy absorbed by the cloud. Modifications occur mainly in low absorption spectral regions, where exchanges between ground and cloudy layers are possible.

Similarly, scattering mainly modifies net exchanges between space and cloudy layers, as shown in \fig{diff_bil_mls_mc}(c), at frequencies where cloudy layers are visible from space. Scattering also brings a number of modifications to net exchanges between space and cloud-free atmospheric layers, in low absorption spectral regions.

Finally, scattering also affects gas-gas net exchanges, as seen on \fig{diff_bil_mls_mc}(d). Net exchanges between cloudy layers and the rest of the atmosphere are affected (because scattering will cause photons emitted into cloudy layers to lose more energy into the clouds, thus reducing net exchanges between these clouds and all other atmospheric layers)~; moreover, net exchanges between cloud-free layers and the rest of the atmosphere are also affected by scattering, mainly because of backscattering by clouds: scattering will cause photons emitted in a given cloud-free layer to be backscattered by optically thick clouds, thus modifying all net-exchanges between the emitting layer and all atmospheric layers involved in the optical paths of the photons.

\section{Summary and conclusions}

The objective of this work is to analyze various simulation results concerning longwave atmospheric radiative transfer. Radiative transfer computations have been carried out using a numerical algorithm based on a Monte-Carlo method developed in previous works \cite{amaury02}, \cite{eymet01}. Radiative transfer results have been expressed in terms of Net Exchange Rates \cite{Green}, allowing a more complete analysis of atmospheric longwave heating rates \cite{Fournier03}, \cite{Fournier04}. Results are presented for a number of configurations already used in a previous work from Fu and al., 1997 \cite{Fu01}.

A first series of results has been presented for clear-sky configurations~; Net Exchange Rate matrices are presented, along with the corresponding longwave atmospheric heating rates. Then radiative budget and its decomposition in terms of net exchanges between atmospheric layers and ground, net exchanges between atmospheric layers and space, and net exchanges between atmospheric layers and the rest of the atmosphere are shown, as a function of frequency and altitude. In a second part, the same results are shown for a series of cloudy atmospheres configurations~; finally, the third part presents the effect of scattering over Net Exchanges Matrices, longwave atmospheric heating rates and radiative budgets for cloudy atmospheres configurations.

In most of the atmosphere the radiative exchanges with boundaries (i.e. ground and space) represent more than 80$\%$ of the heating rate. The exchanges within the atmosphere play a significant role ($\approx 15\%$ to $40\%$ of the heating rate) only in the lower part of the atmosphere, just below the region where the optical thickness (computed from the top of the atmosphere) rises from optically thin to optically thick. The exchanges within the atmosphere may also be important to reduce vertical temperature gradient in the high atmosphere \cite{Bresser} that may appear due to waves and which have not been considered here.

When thick clouds are present, the exchanges with boundaries (the cloud being considered as a boundary too) represent again the higher contribution to the radiative budget. Things are different with the high and thin clouds. They can not be considered as boundaries and radiative exchanges may occur between atmospheric layers on both sides of the cloud.

In both low and high clouds, the effect of scattering by cloud droplets is to widen the photons path length distribution, thus decreasing the cloud total radiative budget. In the meantime, scattering may concentrate radiative net exchanges at cloud edges. Scattering appears to play a more important role for the flux at the top of atmosphere, the flux at the ground and the heating rate inside the cloud. The error on the flux at the top of the atmosphere may be as high as 10$W/m^{2}$ whereas the error on the flux at the ground is less important, except for very dry atmospheres, such as the Arctic profiles, for which both errors are of the same magnitude.

These results and their analysis in terms of Net Exchanges are intended to be used in the future for setting up a parameterization of longwave radiative transfer for representation of the effects of scattering, in a terrestrial Global Circulation Model.

%%%%%%%%%%%%%%%%%%%%%%%%%%%%%%%%la biblio
\bibliographystyle{unsrt}
\bibliography{biblio}

\end{document}